\documentclass[a4paper,12pt]{article}

\usepackage{amsmath, amsthm, amsfonts, amssymb}
\usepackage{graphicx}

\theoremstyle{plain}
\newtheorem{theorem}{Theorem}[section]

\newtheorem{lemma}{Lemma}[section]

\newtheorem{proposition}[lemma]{Proposition}

\theoremstyle{remark}

\def\Om{\Omega}
\def\om{\omega}

\def\g{\gamma}
\def\G{\Gamma}
\def\l{\lambda}
\def\p{\partial}
\def\D{\Delta}

\def\E{\mbox{\rm e}}
\def\a{\alpha}
\def\b{\beta}
\def\si{\sigma}

\def\d{\delta}

\def\z{\zeta}

\def\Odr{\mathcal{O}}
\def\H{W_2}

\def\di{\,\mathrm{d}}

\def\I{\mathrm{I}}

 \DeclareMathOperator{\RE}{Re}
\DeclareMathOperator{\IM}{Im} \DeclareMathOperator{\spec}{\sigma}
\DeclareMathOperator{\discspec}{\sigma_{disc}}
\DeclareMathOperator{\essspec}{\sigma_{ess}}

\DeclareMathOperator{\sgn}{sgn}
\DeclareMathOperator{\tr}{tr}

\numberwithin{equation}{section}

\begin{document}
\allowdisplaybreaks

\title{\textbf{
Distant perturbation asymptotics in window-coupled waveguides. \\
I.~The non-threshold case}}
\author{D.~Borisov$^{a,b}$ and P.~Exner$^{a,c}$}
\date{}
\maketitle

\begin{quote}
{\small {\em a) Nuclear Physics Institute, Academy of Sciences,
25068 \v Re\v z
\\
\phantom{a) }near Prague, Czechia
\\
b) Bashkir State Pedagogical University, October Revolution
St.~3a,
\\
\phantom{a) }450000 Ufa, Russia
\\
c) Doppler Institute for Mathematical Physics and Applied
Mathematics,
\\
\phantom{a) }Czech Technical University, B\v rehov{\'a} 7, 11519
Prague, Czechia }
\\
\phantom{a) }\texttt{borisovdi@yandex.ru},
\texttt{exner@ujf.cas.cz}}
\end{quote}

\begin{quote}
{\small We consider a pair of adjacent quantum waveguides, in
general of different widths, coupled laterally by a pair of
windows in the common boundary, not necessarily of the same
length, at a fixed distance. The Hamiltonian is the respective
Dirichlet Laplacian. We analyze the asymptotic behavior of the
discrete spectrum as the window distance tends to infinity for the
generic case, i.e. for eigenvalues of the corresponding one-window
problems separated from the threshold. }
\end{quote}

\section{Introduction}

Quantum mechanics exhibits various effects which defy our
intuition based on ``classical'' experience. A nice class of
examples are bound states in hard-wall tubes induced solely by
their geometric properties such as bends, protrusions, or
``windows''. Such systems are interesting not only \emph{per se}
but also from the practical point of view as models of various
nanophysical devices, and in a reasonable approximation also of
flat electromagnetic waveguides.

Among numerous questions such models pose an important one
concerns behavior of the spectra in case of two distant
perturbations. One can think of it as of an analogue of the
exponential spectral shift for a pair of distant potential wells,
despite the fact that the usual methods of the Schr\"odinger
operator theory do not work here. The aim of the present paper is
to study this problem in a model example of a pair of laterally
coupled waveguides, or adjacent straight hard-wall strip in the
plane, coupled by a pair of ``windows'' in the common boundary --
we refer to \cite{ESTV}, \cite{BE}, \cite{B1} for a bibliography
concerning such models.

In our recent paper \cite{BE} we dealt with the symmetric
situation where the widths $d_1$, $d_2$ of the two channels were
the same and so were the window widths $a_1, a_2$. The technique
used in these papers employed substantially the fact that the
problem can be decomposed into parts with a definite parity,
which allows one to study a single-window problem with a
perturbation which consists of an additional Dirichlet or
Neumann boundary condition at a segment far from the window.

The approach based on symmetry works no longer if $a_1\ne a_2$.
The man aim of the present work is to demonstrate a different
technique, suitable for the general case, which reduces the
question to analysis of a boundary perturbation at the distant
window. This technique follows the main ideas of \cite{B2}, where
the Dirichlet Laplacian in an $n$-dimensional tube with a pair of
distant perturbations described by two arbitrary operators was
studied. It was assumed in \cite{B2} that these operators are
defined on functions from $\H^2$ vanishing at the boundary, and
this assumption was employed substantially. This is obviously not
true in the problem we study, since the windows enlarge the domain
of the Laplacian beyond the Sobolev space $\H^2$. At the same
time, the general approach of \cite{B2} works in our case with the
appropriate modifications. Moreover, since we restrict ourselves
to the two-dimensional case and specify the nature of the distant
perturbations, we are able to  obtain a more detailed result in
comparison with the general case in \cite{B2}.

In order not to make this study too technical we concentrate in
this paper at the generic case when the ``unperturbed'' energy
is an \emph{isolated} eigenvalue of the one-window problem,
leaving the computationally involved discussion of threshold
resonances to a sequel. The problem will be properly formulated
and the results stated in the next section; the rest of the
paper is devoted to the proofs.

\setcounter{equation}{0}
\section{Statement of the problem and the results}

Let $x=(x_1,x_2)$ be Cartesian coordinates in the plane,
$\Pi^+:=\{x: 0<x_2<\pi\}$ and $\Pi^-:=\{x: -d<x_2<0\}$. With the
natural scaling properties in mind we may suppose without loss of
generality that $d\leqslant\pi$. By $\g_\pm$ we denote two
intervals $\g_\pm:=\{x: |x_1\mp l|<a_\pm,\: x_2=0\}$, from now on
referred to as the \emph{windows}. The numbers $a_\pm$ are assumed
to be fixed throughout the paper while the distance $2l$ between
the windows will be changing playing the role of a large
parameter.

\begin{figure}[t]
\begin{center}
\noindent
\includegraphics[width=12 true cm, height=2.95 true cm]{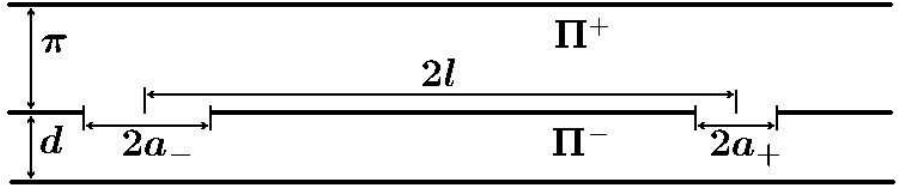}
\end{center}
\caption{Window-coupled waveguides}
\end{figure}

We set $\Pi:=\Pi^+\cup\Pi^-\cup\g_+\cup\g_-$  (cf. Figure 1); the
Hilbert space of our problem is $L_2(\Pi)$. We will employ the
symbol $H$ to denote Friedrichs extension of the negative
Laplacian from the set $C^\infty_0(\Pi)$. We will use the symbols
$\essspec(\cdot)$ and $\discspec(\cdot)$ to indicate the essential
and discrete spectrum, respectively. As we have indicated in the
introduction, this work is devoted to the study of the asymptotic
behavior of isolated eigenvalues of $H$ as $l\to+\infty$. In order
to formulate the main results we have to introduce first some more
notations.

Let $\Omega$ be an open set in $\mathbb{R}^2$ and $\gamma\subset
\overline{\Omega}$. Throughout the paper $W_2^1(\Omega,\gamma)$
will indicate the completion of the set of functions from
$C^\infty( \overline{\Omega})$ having a compact support and
vanishing in the vicinity of the set $\gamma$, taken with
respect to the norm of the Sobolev space $W_2^1(\Omega)$.

We denote $\g_a:=\{x: |x_1|<a, \:x_2=0\}$ so that
$\Pi_a:=\Pi^+\cup\Pi^-\cup\g_a$ is the double waveguide with a
single window centered at $x_1=0$, and $\G_a:=\p\Pi_a$.
Furthermore, we introduce the corresponding cut-off sets
$\Pi_a^b:=\Pi_a\cap\{x: |x_1|<b\}$ and $\G_a^b:=\G_a\cap\{x:
|x_1|<b\}$. Consider the negative Laplacian in $L_2(\Pi_a)$ and
call $H(a)$ its Friedrichs extension in $L_2(\Pi_a)$ from the set
$C_0^\infty(\Pi_a)$ on which it is symmetric; by $\l_m(a),\:
m=1,2,\dots\,$, we denote the isolated eigenvalues of this
operator arranged in the ascending order with the multiplicity
taken into account.

The following results were demonstrated in \cite{ESTV}, \cite{B1}.
\begin{proposition}\label{lm1.1}
For any $a>0$ the essential spectrum of $H(a)$ equals
$[1,+\infty)$ while $\discspec({H}(a))$ is non-empty consisting of
finitely many simple eigenvalues. The eigenfunction associated
with an eigenvalue $\l_m(a)$ has a definite parity: it is even or
odd with respect to $x_1$ if $m$ is odd or even, respectively. In
the particular case $d=\pi$ the eigenfunctions are even in the
variable $x_2$.
\end{proposition}

The eigenfunctions associated with the eigenvalues $\l_n(a)$ will
be denoted as $\psi_n(\cdot,a)$ and assumed to be normalized, i.e.
to be unit vectors in $L_2(\Pi_a)$. It is easy to check that
$\psi_n(\cdot,a)\in C^\infty(\Pi_a)$. We put $\sigma_*:=
\discspec{H(a_+)}\cup\discspec{H(a_-)}$; an element $\l_*\in
\sigma_*$ of this set will be called \emph{simple} if $\l_*$
belongs to one of the sets $\discspec(H(a_\pm))$ only and
\emph{double} otherwise. Furthermore, we set $\boldsymbol{a}:=
(a_+,a_-)$.

\smallskip

With these preliminaries we can formulate the first main result of
this paper.
\begin{theorem}\label{th1.0}
For any $l>0$, $a_\pm>0$ the operator $H$ has the essential
spectrum equal to $[1,+\infty)$ and finitely many isolated
eigenvalues. The number of the isolated eigenvalues of $H$ is
independent of the window distance provided
$l\geqslant\max\{a_-,a_+\}$. In the limit $l\to+\infty$ each
isolated eigenvalue of the operator $H$ converges to one of the
numbers from the set $\sigma_*$ or to the threshold of
$\essspec(H)$.
\end{theorem}

By $\Xi_a$ we indicate the set of all bounded domains
$S\subset\Pi_a$ having smooth boundary and separated from the
edges of $\g_a$ by a positive distance; we stress that the case
$\p S\cap \p\Pi_a\not=\emptyset$ is not excluded. For any $\l$
such that $\RE\l\leqslant 1$ we denote
$\kappa_1^+=\kappa_1^+(\l):= \sqrt{1-\l}$ and
$\kappa_1^-=\kappa_1^-(\l):= \sqrt{\frac{\pi^2}{d^2}-\l}$, where
the branch of the root is specified by the requirement that the
functions are analytic in $\mathbb{S}_\d$ and $\sqrt{1}=1$.

The following statements will be proven in Section~\ref{s:
onewin}.

\begin{proposition}\label{lm1.2}
In the limit $x_1\to\pm\infty$ the eigenfunction $\psi_n$ of
$H(a)$ behaves as
\begin{equation}\label{1.4}
\begin{aligned}
&\psi_n(x,a)=(\pm
1)^{n+1}c(\l_n,a)\E^{-\kappa_1^+(\l_n)|x_1|}\sin
x_2+\Odr(\E^{-\RE\sqrt{4-\l}|x_1|}),&& x_2\in[0,\pi],
\\
&\psi_n(x,a)=\Odr(\E^{-\RE \kappa_1^-(\l)|x_1|}), &&
x_2\in[-d,0],
\end{aligned}
\end{equation}
if $d<\pi$, and
\begin{equation}\label{1.4a}
\begin{aligned}
&\psi_n(x,a)=(\pm
1)^{n+1}c(\l_n,a)\E^{-\kappa_1^+(\l_n)|x_1|}\sin
|x_2|+\Odr(\E^{-\RE\sqrt{4-\l}|x_1|}),&& x_2\in[-\pi,\pi],
\end{aligned}
\end{equation}
in the case of equal-width channels, $d=\pi$. In these relations
\begin{equation}\label{1.5}
c(\l_n,a)=\frac{1}{\pi
\kappa_1^+(\l_n)}\int\limits_{\g_a}\psi_n(x,a)\E^{\kappa_1^+(\l_n)x_1}\di
x_1=\frac{(-1)^{n+1}}{\pi
\kappa_1^+(\l_n)}\int\limits_{\g_a}\psi_n(x,a)\E^{-\kappa_1^+(\l_n)x_1}\di
x_1,
\end{equation}
and $c(\l_i,a)\not=0$, $i=1,2$. The asymptotic relations
(\ref{1.4}), (\ref{1.4a}) give rise to valid formul{\ae} when
both their sides are differentiated.
\end{proposition}

\begin{proposition}\label{lm1.3}
For any $\l\in(-\infty,1)\setminus\discspec(H(a))$ there exists
a unique solution of the boundary value problem
\begin{equation}\label{1.3a}
\begin{gathered}
(\D+\l)U=0,\quad x\in\Pi_a\setminus\g_a,\qquad U=0,\quad
x\in\p\Pi_a,
\\
\frac{\p U}{\p x_2}\Big|_{x_2=+0}- \frac{\p U}{\p
x_2}\Big|_{x_2=-0}=\E^{\kappa_1^+(\l)x_1}, \quad x\in\g_a,
\end{gathered}
\end{equation}
belonging to $\H^1(\Pi_a)$. For large values of $|x_1|$ this
function is infinitely differentiable and in the limit
$x_1\to+\infty$ it behaves as
\begin{equation}\label{1.3b}
\begin{aligned}
&U(x,\l,a)=c(\l,a)\E^{-\kappa_1^+(\l_n)x_1}\sin
x_2+\Odr(\E^{-\RE\sqrt{4-\l}x_1}),&& x_2\in[0,\pi],
\\
&U(x,\l,a)=\Odr(\E^{-\RE \kappa_1^-(\l)x_1}), && x_2\in[-d,0],
\end{aligned}
\end{equation}
if $d<\pi$, and
\begin{equation}\label{1.3c}
\begin{aligned}
&U(x,\l,a)=c(\l,a)\E^{-\kappa_1^+(\l_n)x_1}\sin
|x_2|+\Odr(\E^{-\RE\sqrt{4-\l}x_1}),&& x_2\in[-\pi,\pi],
\end{aligned}
\end{equation}
in the case $d=\pi$, where the coefficient is given by
\begin{equation}\label{1.10}
c(\l,a)=\frac{1}{\pi
\kappa_1^+(\l)}\int\limits_{\g_a}U(x,\l,a)\E^{\kappa_1^+(\l)x_1}\di
x_1.
\end{equation}
This coefficient is negative for $\l<\l_1(a)$. The asymptotic
relations (\ref{1.3b}), (\ref{1.3c}) give rise to valid
formul{\ae} when both their sides are differentiated.
\end{proposition}

For the double window $\g_+\cup\g_-$ we indicate by $\Xi$ the
set of all bounded domains $S\subset\Pi$ having smooth boundary
and separated from the edges of $\g_\pm$ by a positive distance;
the case $\p S\cap \p\Pi\not=\emptyset$ is again not excluded.
For brevity we will introduce a two-valued symbol, $\tau:=1$ if
$d<\pi$ and $\tau:=2$ if $d=\pi$.

Continuing the list of the main results we make the following
claims.

\begin{theorem}\label{th1.2}
Suppose that $\l_*\in\sigma_*$ is simple being an eigenvalue
$\l_n(a_\pm)$ of the operator $H(a_\pm)$. Then there is a unique
eigenvalue of the operator $H$ converging to $\l_*$ as
$l\to+\infty$. This eigenvalue is simple and behaves
asymptotically as follows,
\begin{align}
&\l^\pm(l,\boldsymbol{a})=\l_*+\mu^\pm(l,\boldsymbol{a})
\E^{-4\kappa_1^+(\l_*)l}+\Odr(l^2\E^{-8\kappa_1^+(\l_*)l}+
\E^{-2l(\kappa_1^+(\l_*)+\rho)}),\label{1.11}
\\
&
\begin{aligned}
&\mu^\pm(l,\boldsymbol{a}):=\tau\pi c(\l_*,a_\pm)c^2(\l_*,a_\mp)
\kappa_1^+(\l_*),
\end{aligned}\label{1.12}
\end{align}
where $\rho=\rho(\l):=\min\{\kappa_1^-(\l),\sqrt{4-\l}\}$ if
$d<\pi$, and $\rho=\rho(\l):=\sqrt{4-\l}$ if $d=\pi$. The
associated eigenfunction $\psi^\pm(x,l,\boldsymbol{a})$
satisfies the relation
\begin{equation}\label{1.13}
\psi^\pm(x,l,\boldsymbol{a})=\psi_n(x_1\mp
l,x_2,a_\pm)+\Odr(\E^{-2\kappa_1^+(\l_*)l})
\end{equation}
in the norms of both the $\H^1(\Pi)$ and $\H^2(S)$ for each
$S\in\Xi$.
\end{theorem}

\begin{theorem}\label{th1.3}
Suppose that $\l_*\in\sigma_*$ is double and $\l_*=\l_n(a_-)=
\l_m(a_+)$. Then there exist either two simple eigenvalues
$\l^\pm(l,\boldsymbol{a})$ or one double eigenvalue
$\l^-(l,\boldsymbol{a})=\l^+(l,\boldsymbol{a})$ of the operator
$H$ converging to $\l_*$ as $l\to+\infty$. The asymptotic
expansions of  these eigenvalues read as follows,
\begin{align}
&\l^\pm(l,\boldsymbol{a})=\l_*\pm|\mu(l,\boldsymbol{a})|
\E^{-2\kappa_1^+(\l_*)l}+
\Odr(l\E^{-4\kappa_1^+(\l_*)l}+\E^{-2\rho(\l_*)l}),\label{1.14}
\\
&
\begin{aligned}
&\mu(l,\boldsymbol{a})=(-1)^{m+1}\tau\pi
\kappa_1^+(\l_*)c(\l_*,a_-)c(\l_*,a_+),
\end{aligned}\label{1.15}
\end{align}
\end{theorem}

\begin{theorem}\label{th1.4}
Suppose the hypothesis of Theorem~\ref{th1.3} holds true. If
$\mu(l,\boldsymbol{a})\not=0$, the eigenvalues
$\l^+(l,\boldsymbol{a})$ $\l^-(l,\boldsymbol{a})$ do not
coincide and are simple. The associated eigenfunctions
$\psi^\pm(x,l,\boldsymbol{a})$ satisfy the relations
\begin{equation}\label{1.16}
\psi^\pm(x,l,\boldsymbol{a})=\psi_n(x_1+l,x_2,a_-)\mp
\psi_m(x_1-l,x_2,a_+)
\sgn\mu(l,\boldsymbol{a})+\Odr(\E^{-2\kappa_1^+(\l_*)l})
\end{equation}
in the norms of $\H^1(\Pi)$ and $\H^2(S)$ for each $S\in\Xi$. If
$\l^-(l,\boldsymbol{a})=\l^+(l,\boldsymbol{a})$ is a double
eigenvalue, the associated eigenfunctions $\psi^\pm(x,l,
\boldsymbol{a})$ satisfy the relations
\begin{equation}\label{1.17}
\begin{aligned}
&\psi^+(x,l,\boldsymbol{a})=\psi_n(x_1+l,x_2,a_-)
+\Odr(\E^{-2\kappa_1^+(\l_*)l}),
\\
&\psi^-(x,l,\boldsymbol{a})=\psi_m(x_1-l,x_2,a_-)
+\Odr(\E^{-2\kappa_1^+(\l_*)l}),
\end{aligned}
\end{equation}
in the norm of $\H^1(\Pi)$ and $\H^2(S)$ for each $S\in\Xi$.
Finally, if $\mu(l,\boldsymbol{a})=0$ and
$\l^-(l,\boldsymbol{a}) \not=\l^+ (l,\boldsymbol{a})$, the
eigenvalues $\l^\pm(l, \boldsymbol{a})$ are simple and the
associated eigenfunctions satisfy the relations
\begin{equation}\label{1.18}
\psi^\pm(x,l,\boldsymbol{a})=c_+^\pm\psi_n(x_1+l,x_2,a_-)+c_-^\pm
\psi_m(x_1-l,x_2,a_+)+\Odr(\E^{-2\kappa_1^+(\l_*)l}),
\end{equation}
where the vectors $\boldsymbol{c}^\pm:=\left(
\begin{smallmatrix}
c_+^\pm
\\
c_-^\pm
\end{smallmatrix}
\right)$ are nontrivial solutions to the system (\ref{4.16})
with $\l=\l^\pm$ such that
$\|\boldsymbol{c}^\pm\|_{\mathbb{R}^2}=1$.
\end{theorem}

The leading terms of the asymptotics (\ref{1.11}), (\ref{1.14})
are non-zero provided the corresponding coefficients
$c(\l_*,a_\pm)$ are non-zero. We know from
Propositions~\ref{lm1.2}, \ref{lm1.3} that this is true at least
for $c(\l,a)$ as $\l\leqslant\l_1(a)$ or $\l=\l_2(a)$. For
instance, if $\l_1(a_-)<\l_1(a_+)$, the eigenvalue of the operator
$H$ converging to $\l_1(a_-)$ has the asymptotic expansion
(\ref{1.11}), and the coefficient (\ref{1.12}) of leading term is
non-zero. Moreover, due to Proposition~\ref{lm1.3} this
coefficient is negative. If $a_\pm$ are such that
$\l_1(a_-)=\l_2(a_+)$, the eigenvalues of the operator $H$
converging to $\l_*=\l_1(a_-)=\l_2(a_+)$ have the asymptotics
expansions (\ref{1.14}), and the coefficients of the leading terms
are non-zero. By Theorem~\ref{th1.4} the ``perturbed'' eigenvalues
are simple and the associated eigenfunctions satisfy the
identities (\ref{1.16}) in this case. We also stress that in this
case the leading terms of the asymptotic expansions (\ref{1.14})
have the same modulus but different signs. This phenomenon is
known in double-well problems with symmetric wells. It also occurs
in the symmetric case, $a_-=a_+$ and $d_-=d_+$, as we have shown
in \cite{BE}.

We conjecture that the coefficient $c(\l,a)$ is non-zero for all
values $a$ and $\l<1$. If it is true, this fact would imply that
the leading terms in the asymptotics (\ref{1.11}), (\ref{1.14})
are non-zero. In turn, this fact together with
Theorem~\ref{th1.4} would imply that a double $\l_*\in\si_*$
splits into two simple ''perturbed'' eigenvalues and the
formul{\ae} (\ref{1.16}) are valid for the associated
eigenfunctions.

\section{Analysis of the one-window problem} \label{s: onewin}

In this section we shall study the following boundary value
problem,
\begin{equation}\label{3.11}
\begin{gathered}
(\D+\l)u=0,\quad x\in\Pi_a\setminus\g_a, \qquad u=0,\quad
x\in\G_a,
\\
\frac{\p u}{\p x_2}\Big|_{x_2=+0}- \frac{\p u}{\p
x_2}\Big|_{x_2=-0}=f, \quad x\in\g_a.
\end{gathered}
\end{equation}
The function $f$ is assumed to be an element of $L_2(\g_a)$. A
solution to this problem is understood in a generalized sense,
more specifically, as a function belonging to
$\H^1(\Pi_a^b,\G_a^b)$ for each $b>0$ and satisfying the equation
\begin{equation}\label{3.12}
-(\nabla u,\nabla \z)_{L_2(\Pi_a)}+\l(u,\z)_{L_2(\Pi_a)}-
(f,\z)_{L_2(\g_a)}=0
\end{equation}
for any $\z\in C_0^\infty(\Pi_a)$. By standard
smoothness-improving results about solutions to elliptic boundary
value problems, cf.~\cite[Ch. 4, \S 2]{Mh}, the said solution
belongs to $C^\infty(\overline{\Pi^+}\cup\overline{\Pi^-}\setminus
\overline{\g}_a)$. As we have said in the introduction we will
deal in this paper with the non-threshold case only. Thus the
parameter $\l$ is supposed to belong to $\mathbb{S}_\d$ for a
fixed $\d>0$, where $\mathbb{S}_\d$ is a set of all $\l$ separated
from the halfline $[1,+\infty)$ by a distance not less than $\d$.

We seek a solution to the problem (\ref{3.11}) belonging to
$L_2(\Pi_a)$. We fix $\b>a$ and put $P:=\{x: |x_1|<a+\b,\,
0<x_2<d_0\}$. The number $d_0$ here is chosen so that $d_0<d$, and
the lowest eigenvalue of the negative Laplacian in $P$ subject to
Dirichlet boundary condition on $\p P\setminus \overline{\g}_a$
and to Neumann one on $\g_a$ exceeds two. We consider the boundary
value problem
\begin{equation}\label{3.13}
(\D+\l)  \widetilde{u}=0,\quad x\in P,\qquad
\widetilde{u}=0,\quad x\in\p P,\qquad \frac{\p\widetilde{u}}{\p
x_2}=\frac{1}{2}f,\quad x\in\g_a,
\end{equation}
which is again treated in the weak sense,
\begin{equation}\label{3.14}
-(\nabla \widetilde{u},\nabla\z)_{L_2(P)}
+\l(\widetilde{u},\z)_{L_2(P)}-\frac{1}{2}(f,\z)_{L_2(\g_a)}=0
\end{equation}
for each function $\z\in C^\infty(P)$ vanishing in a
neighborhood of $\p P\setminus\overline{\g}_a$. The problem
(\ref{3.13}) is uniquely solvable in the space $\H^1(P,\p
P\setminus\g_a)$ and the solution belongs to
$C^\infty(\overline{P}\setminus\overline{\g}_a)$  -- see
\cite[Chap. I\!I, \S 5, Rem. 5.1]{Ld} and \cite[Chap. I\!V, \S
2]{Mh}).

Let $\chi_1=\chi_1(x)$ be an infinitely differentiable function,
even w.r.t. the variable $x_2$, equal to one if $|x_1|<a+\b/6$ and
$|x_2|<d_0/6$, and vanishing for $|x_1|>a+\b/3$ or $|x_2|>d_0/3$.
We extend the function $\widetilde{u}$ in an even way for $x_2<0$
setting $\widetilde{u}(x):=\widetilde{u}(x_1,-x_2)$ as $x_2<0$ and
denote $u_f(x):=\chi_1(x)\widetilde{u}(x)$.

\begin{lemma}\label{lm3.5}
The function $u_f$ belongs to $\H^1(\Pi_a,\G_a)\cap
C^\infty(\Pi_a)$ and satisfies the equation
\begin{equation}\label{3.15}
-(\nabla u_f,\nabla \z)_{L_2(\Pi_a)}+\l(u_f,\z)_{L_2(\Pi_a)}
-(f,\z)_{L_2(\g_a)}=(F,\z)_{L_2(\Pi_a)}
\end{equation}
for any $\z\in C_0^\infty(\Pi_a)$, where
\begin{equation*}
F=T_{1}(\l,a)f:=2\nabla\widetilde{u}\cdot\nabla\chi_1+
\widetilde{u}\D\chi_1\,.
\end{equation*}
The operator $T_{1}: L_2(\g_a)\to L_2(\{x: |x_1|<a+\b/3,
|x_2|<d_0/3\})$ is linear, bounded, and holomorphic in
$\lambda$. The operator $T_{2}(\l,a)f:=u_f$ is linear, bounded,
and holomorphic in $\l$ as a map from $L_2(\g_a)$ into
$\H^1(\Pi_a,\G_a)$, $\H^2(S)$, and
$\H^2(\Pi^\pm\setminus\Pi_a^\b)$, where $S\in\Xi_a$ is such that
$S\subset\Pi^+$ or $S\subset\Pi^-$.
\end{lemma}
\begin{proof}
Let $\{\widetilde{u}^{(j)}\}$ be a sequence of functions from
$C^\infty(\overline{P})$ vanishing in a neighborhood of $\p
P\setminus\overline{\g}_a$, which converges to $\widetilde{u}$
in $\H^1(P)$. It is easy to see that the functions
$u^{(j)}_f(x):=\chi_1(x)\widetilde{u}^{(j)}(x)$ belong to
$\H^1(\Pi_a,\G_a)$, and that they converge to $w_f$ in the norm
of $\H^1(\Pi_a)$ as $j\to\infty$, so $u_f\in\H^1(\Pi_a,\G_a)$.
Next we observe that $u_f$ belongs to $C^\infty(\Pi_a)$ as it
follows from the fact that $\widetilde{u}\in
C^\infty(\overline{P}^+ \setminus\overline{\g}_a)$. Since the
function $u_f$ is even in the variable $x_2$, we find that for
each $\z\in C_0^\infty(\Pi_a)$ the left-hand side of
(\ref{3.15}) equals twice the expression
\begin{equation*}
-(\nabla u_f,\nabla\z^+)_{L_2(P)}+\l (u_f,\z^+)_{L_2(P)}-
(f,\z^+)_{L_2(\g_a)}\,,
\end{equation*}
where $\z^+(x):=\z(x)+\z(x_1,-x_2)$. In view of (\ref{3.14}) and
the definition of $\chi_1$ we get
\begin{align*}
&-(\nabla u_f,\nabla\z)_{L_2(\Pi_a)}+\l (u_f,\z)_{L_2(\Pi_a)}-
(f,\z)_{L_2(\g_a)}
\\
=&2\Bigg(-(\nabla\widetilde{u},\nabla(\chi_1\z^+))_{L_2(P)}+\l
(u_f,\chi_1\z^+)_{L_2(P)}-\frac{1}{2} (f,\chi_1\z^+)_{L_2(\g_a)}
\\
&+(\nabla\widetilde{u},\z^+\nabla\chi_1)_{L_2(P)}-(
\widetilde{u}\nabla\chi_1, \nabla\z^+)_{L_2(P)}\Bigg)
\\
=&2\Big((\nabla
\widetilde{u},\z^+\nabla\chi_1)_{L_2(P)}-(\widetilde{u}\nabla\chi_1,
\nabla\z^+)_{L_2(P)}\Big)
\\
=&2\Big((\nabla\widetilde{u},\z^+\nabla\chi_1)_{L_2(P)}
+(\mathrm{div}\,\widetilde{u}\nabla\chi_1,
\z^+)_{L_2(P)}\Big)=(F,\z)_{L_2(\Pi_a)}.
\end{align*}
The boundedness of the operator $T_{1}(\l,a)$ follows from the
above mentioned theorems on improving smoothness of solutions to
elliptic boundary value problems.

In order to check that $T_{1}$ is holomorphic in the variable
$\l$ we just need to show that the mapping $f\mapsto
\widetilde{u}$ is bounded and holomorphic as an operator family
from $L_2(\g_a)$ into $\H^1(P)$ and $\H^2(S\cap P)$, where
$S\in\Xi_a$ and $S\subset\overline{\Pi^+}$. To prove the last
claim it is sufficient to reduce the boundary value problem to
an operator equation in $\H^1(P,\p P\setminus\overline{\g}_a)$
in the standard way -- see \cite[Ch. I\!I, \S 2]{Mh} -- and to
apply then Proposition 4.5 of \cite[Ch. X\!I, \S 4]{SP}.
\end{proof}

We seek the solution to the problem (\ref{3.11}) in the form
$u=u_f+\widehat{u}$. As $u_f$ is compactly supported, the function
$\widehat{u}$ has to be an element of $L_2(\Pi_a)$. It follows
from (\ref{3.12}), (\ref{3.15}) that the function $\widehat{u}$
must also obey the integral relation
\begin{equation}\label{3.24}
-(\nabla \widehat{u},\nabla\z)_{L_2(\Pi_a)}+
\l(\widehat{u},\z)_{L_2(\Pi_a)}=-(F,\z)_{L_2(\Pi_a)}
\end{equation}
for any $\z\in C_0^\infty(\Pi_a)$. Thus $\widehat{u}$ has to solve
the boundary value problem
\begin{equation}\label{3.25}
-(\D+\l)\widehat{u}=F,\quad x\in\Pi_a, \qquad u=0,\quad
x\in\G_a,
\end{equation}
belonging to $L_2(\Pi_a)$ and $\H^1(\Pi_a^b,\G_a^b)$ for each
$b>0$. By Theorem~4.6.8 of \cite[Ch. 4, \S 4.6]{BEH} any solution
of this problem belonging to $L_2(\Pi_a)$ is an element of the
operator domain of $H(a)$. In this way the problem (\ref{3.25})
can be cast into the form $(H(a)-\l)\widehat{u}=F$, which in turn
gives $\widehat{u}=\big(H(a)-\l\big)^{-1}F$.

Let us next denote $T_{3}(\l,a):=T_{2}(\l,a) +(H(a)-\l)^{-1}
T_{1}(\l,a)$. In order to analyze properties of this operator we
need an additional notation and a lemma. For any numbers
$b_1,b_2,b_3\in \mathbb{R}$ we set $\Om_\pm:=\{x: \pm x_1>b_1,\,
b_2<x_2<b_3\}$ and $\om_\pm:=\{x: \pm x_1>b_1\}\cap\p\Om_\pm$.

\begin{lemma}\label{lm3.12}
Let $v\in\H^1(\Om_\pm)$ be a solution to the problem
\begin{equation*}
(\D+\l)v=0,\quad x\in\Om^\pm,\qquad v=0,\quad x\in\om^\pm,
\end{equation*}
and $\,0<b_2-b_3\leqslant \pi$, $\l\in\mathbb{S}_\d$. Then the
function $v$ can be represented as
\begin{equation}\label{3.40}
v(x,\l)=\sum\limits_{j=1}^{\infty}\a_j(\l)
\exp\left(-\sqrt{\frac{\pi^2 j^2}{(b_3-b_2)^2}-\l}\:(\pm
x_1-b_1)\right) \sin\frac{\pi j}{b_3-b_2}(x_2-b_2),
\end{equation}
where
\begin{equation}\label{3.41}
\a_j(\l):=\frac{2}{b_3-b_2}\int\limits_{b_2}^{b_3} v(a_1,x_2,\l)
\sin\frac{\pi j}{b_3-b_2}(x_2-b_2)\di x_2.
\end{equation}
The series (\ref{3.40}) converges in the norms of $\H^m(\{x: \pm
x_1>b_4,b_2<x_2<b_3\})$, $m\geqslant0$, for any $b_4>b_1$. The
coefficients $\a_j$ satisfy the condition
\begin{equation}\label{3.42}
\frac{\pi}{2}\sum\limits_{j=1}^{\infty} |\a_j|^2=\|v(\pm
b_1,\cdot,\l)\|_{L_2(b_2,b_3)}.
\end{equation}
\end{lemma}
This lemma is a particular case of Lemma~3.3 of \cite{B2} so we
skip the proof.

\begin{lemma}\label{lm3.13}
The operator $T_{3}(\l,a)$ is bounded and meromorphic in $\l\in
\mathbb{S}_\d$ as a map from $L_2(\g_a)$ into $\H^1(\Pi_a,\G_a)$
and into $\H^2(S)$ for each $S\in\Xi_a$. Its poles coincide with
the eigenvalues of the operator $H(a)$. For any $\l$ close to an
eigenvalue $\l_n$ of $H(a)$ the representation
\begin{equation}
T_{3}(\l,a)=
\frac{\psi_n}{\l-\l_n}T_{4}(a)+T_{5}(\l,a)\label{3.9}
\end{equation}
holds true. Here $T_{4}(a)f:=(f,\psi_n)_{L_2(\g_a)}$ and the
operator $T_{5}$ is bounded and holomorphic in $\l\in
\mathbb{S}_\d$ as a map from $L_2(\g_a)$ into
$\H^1(\Pi_a,\G_a)$. The operator $T_{5}$ is also bounded and
holomorphic as a map into $\H^2(S)$ for each $S\in\Xi_a$.
\end{lemma}
\begin{proof}
In accordance with \cite[Ch. 5, \S 3.5]{K} the operator
$(H(a)-\l)^{-1}$ is bounded and meromorphic in $L_2(\Pi_a)$, its
poles coincide with the eigenvalues of $H(a)$ and for $\l$ close
to $\l_n$ the representation
\begin{equation}\label{3.27}
\big(H(a)-\l\big)^{-1}=-\frac{\psi_n}{\l-\l_n}
(\cdot,\psi_n)_{L_2(\Pi_a)}+T_{6}(\l,a)
\end{equation}
is valid, where the operator $T_{6}(\l,a)$ is bounded and
holomorphic in $\l$ in the vicinity of $\l_n$. The function
$\check{u}:=T_{6}(\l,a)F$ is a solution to the boundary value
problem (\ref{3.25}) with $F$ replaced by
$\check{F}:=F-(F,\psi_n)_{L_2(\Pi_a)}\psi_n$; it means that
\begin{equation*}
\|\nabla\check{u}\|_{L_2(\Pi_a)}^2-\l\|\check{u}\|_{L_2(\Pi_a)}^2=
(\check{F},\check{u})_{L_2(\Pi_a)}.
\end{equation*}
This relation together with (\ref{3.27}) imply that the operator
$T_{6}$ is bounded and holomorphic as a map into $\H^1(\Pi_a)$
as well. Using again the smoothness-improving theorems mentioned
above we conclude that the operator $T_{6}$ is also bounded and
holomorphic in $\lambda$ as a map into $\H^2(S)$ for each
$S\in\Xi_a$.

Since the function $\psi_n$ is an element of $\H^1(\Pi_a,\G_a)$,
the relation (\ref{3.15}) is valid for $\z=\psi_n$. For any
$f\in L_2(\g_a)$ the function $T_{1}(\l,a)f$ is compactly
supported, hence we have
\begin{equation*}
(T_{1}(\l_n,a)f,\psi_n)_{L_2(\Pi_a)}=-(\nabla u_f,\nabla
\psi_n)_{L_2(\Pi_a)}+\l(u_f,\psi_n)_{L_2(\Pi_a)}-
(f,\psi_n)_{L_2(\g_a)}.
\end{equation*}
According to Lemma~\ref{lm3.5}, the function $u_f$ belongs to
$\H^1(\Pi_a,\G_a)$, which allows us to proceed with the
calculations,
\begin{gather*}
-(\nabla u_f,\nabla
\psi_n)_{L_2(\Pi_a)}+\l(u_f,\psi_n)_{L_2(\Pi_a)}=0,
\\
(T_{1}(\l_n,a)f,\psi_n)_{L_2(\Pi_a)}=-(f,\psi_n)_{L_2(\g_a)}=-T_{4}f.
\end{gather*}
Substituting the relation thus obtained together with
(\ref{3.27}) into the definition of the operator $T_{3}$ and
taking into account Lemma~\ref{lm3.5}, we arrive finally at the
statement of the lemma.
\end{proof}

Let us next fix a number $\widetilde{a}>0$. For any
$l\geqslant(a+\widetilde{a})$ we define operators
$T_{7}^\pm(\l,l,a,\widetilde{a})$ which map an arbitrary $v\in
\H^1(\Pi_a^a)$ into the function
\begin{align}
&(T_{7}^\pm v)(x_1,\l,l):=\sum\limits_{j=1}^{\infty} j\a_j^\pm
\E^{\mp \kappa_j^+(\l)(x_1\mp a)}\E^{-2\kappa_j^+(\l)l}-
\sum\limits_{j=1}^{\infty} \frac{\pi j}{d}\b_j^\pm \E^{\mp
\kappa_j^-(\l)(x_1\mp a)}\E^{-2\kappa_j^-(\l)l},\nonumber
\\
&\a_j^\pm=\frac{2}{\pi}\int\limits_0^\pi v(a,x_2)\sin jx_2\di
x_2,\quad \b_j^\pm=\frac{2}{d}\int\limits_{-d}^0 v(a,x_2)\sin
\frac{\pi j}{d}x_2\di x_2,\nonumber
\\
&\kappa_j^+(\l):=\sqrt{j^2-\l},\quad
\kappa_j^-(\l):=\sqrt{\frac{\pi^2 j^2}{d^2}-\l},\quad j\geqslant
2.\label{3.35a}
\end{align}
The branch of the root in the definition of the functions
$\kappa_j$ is specified by the requirement that the functions
are analytic in $\mathbb{S}_\d$ and $\sqrt{1}=1$.

\begin{lemma}\label{lm3.7}
The operators $T_{7}^\pm: \H^1(\Pi_a^a)\to
L_2(\g_{\widetilde{a}})$ are well defined, bounded and
holomorphic in $\l\in \mathbb{S}_\d$. The estimates
\begin{equation*}
\left\|\frac{\p^i T_{7}^\pm}{\p\l^i}\right\|\leqslant
Cl^i\E^{-(2l-a-\widetilde{a})\RE \kappa_1^+(\l)},\quad i=0,1,2,
\end{equation*}
hold true uniformly w.r.t. $\l\in \mathbb{S}_\d$ and $l\geqslant
(a+\widetilde{a})$.
\end{lemma}
\begin{proof}
We will prove the lemma for $T_{7}^+$ only, the argument for
$T_{7}^-$ is similar. The function $u$ belongs to
$\H^1(\Pi_a^a)$, hence we have the estimate
\begin{equation*}
\sum\limits_{j=1}^{\infty}\big(|\a_j^\pm|^2+|\b_j^\pm|^2\big)
\leqslant C\|u\|_{\H^1(\Pi_a)}^2,
\end{equation*}
where the constant $C$ is independent of $\l\in \mathbb{S}_\d$ and
$l\geqslant (a+\widetilde{a})$. Employing this inequality we infer
that
\begin{align*}
&\left\|\sum\limits_{j=1}^{\infty} j\a_j^+ \E^{-
\kappa_j^+(\l)(\cdot-
a)}\E^{-2\kappa_j^+(\l)l}\right\|_{L_2(\g_{\widetilde{a}})}
\leqslant \sum\limits_{j=1}^{\infty} j|\a_j| \E^{-2l\RE
\kappa_j^+(\l) }
\|\E^{-\kappa_j^+(\l)(\cdot-a)}\|_{L_2(-\widetilde{a},\widetilde{a})}
\\
&\leqslant C\sum\limits_{j=1}^{\infty}
\frac{j|\a_j^+|}{\sqrt{\RE
\kappa_j^+(\l)}}\E^{-(2l-a-\widetilde{a})\RE \kappa_j^+(\l)}
\\
&\leqslant C\left(\sum\limits_{j=1}^{\infty}
|\a_j|^2\right)^{1/2}\left(\sum\limits_{j=1}^{\infty}
\frac{j^2\E^{-2(2l-a-\widetilde{a})\RE
\kappa_j^+(\l)}}{|\kappa_j^+(\l)|}\right)^{1/2}
\\
&\leqslant C\|v\|_{\H^1(\Pi_a^a)}\E^{-(2l-a-\widetilde{a}) \RE
\kappa_1^+(\l)} \left(\sum\limits_{j=1}^{\infty}
\frac{j^2\exp\left(-2(2l-a-\widetilde{a})\RE
(\kappa_j^+(\l)-\kappa_1^+(\l))\right)}{\sqrt{|\IM\l|}}\right)^{1/2}
\\
&\leqslant C\E^{-(2l-a-\widetilde{a})}\|v\|_{\H^1(\Pi_a^a)},
\end{align*}
where $C$ is independent of $\l\in \mathbb{S}_\d$. In the same way
one can prove that
\begin{align*}
&\left\|\sum\limits_{j=1}^{\infty} \frac{\pi j}{d}\b_j^+ \E^{-
\kappa_j^-(\l)(\cdot-
a)}\E^{-2\kappa_j^-(\l)l}\right\|_{L_2(\g_{\widetilde{a}})}
\leqslant C\E^{-(2l-a-\widetilde{a})\RE
\kappa_1^-(\l)}\|v\|_{\H^1(\Pi_a^a)},
\end{align*}
The last two estimates imply that the operator
$T_{7}^+:\H^1(\Pi_a^a)\to L_2(\g_{\widetilde{a}})$ is well
defined and bounded. One can check easily that
\begin{align*}
\left(\frac{\p T_{7}^+
v}{\p\l}\right)(x_1,\l,l):=&\sum\limits_{j=1}^{\infty}
\frac{j\a_j^+}{2\kappa_j^+(\l)}(x_1-a+2l)\E^{-\kappa_j^+(\l)(x_1-
a)}\E^{-2\kappa_j^+(\l)l}
\\
&-\sum\limits_{j=1}^{\infty} \frac{\pi j
\b_j^+(x_1-a+2l)}{2\kappa_j^-(\l)d}\E^{-\kappa_j^-(\l)(x_1-
a)}\E^{-2\kappa_j^-(\l)l}.
\end{align*}
Repeating the argument which yielded the estimate for $T_{7}^+v$
we can establish that
\begin{align*}
&\left\|\frac{\p T_{7}^\pm
v}{\p\l}\right\|_{L_2(\g_{\widetilde{a}})} \leqslant
Cl\E^{-(2l-a-\widetilde{a})\RE
\kappa_1^-(\l)}\|v\|_{\H^1(\Pi_a^a)}
\end{align*}
with the constant $C$ independent of $\l\in \mathbb{S}_\d$ and
$l\geqslant (a+\widetilde{a})$. Consequently, the operator
$\frac{\p T_{7}^+}{\p\l}$ exists, it is bounded and the stated
estimate for its norm holds true. The norm estimate for
$\frac{\p^2 T_{7}^+}{\p\l^2}$ is obtained in a similar way.
\end{proof}

For any $l\geqslant (a+\widetilde{a})$ we define operators
$T_{8}^\pm(\l,l,a,\widetilde{a})$ which map any $f\in L_2(\g_a)$
into the function
\begin{equation*}
\frac{\p u}{\p x_2}(x_1\pm 2l,+0,\l)-\frac{\p u}{\p x_2}(x_1\pm
2l,-0,\l),\quad x_1\in(-\widetilde{a},\widetilde{a}).
\end{equation*}
Here $u$ is a solution to the boundary value problem
(\ref{3.11}) belonging to $\H^1(\Pi)$. Taking into account
Lemma~\ref{lm3.13} together with the boundedness of the
embedding $\H^1(\Pi_a)$ into $L_2(\{x: |x_1\pm
2l|<\widetilde{a}, x_2=0\})$, we conclude that the operators
$T_{8}^\pm: L_2(\g_a)\to L_2(\g_{\widetilde{a}})$ are bounded
and holomorphic in $\l\in\mathbb{S}_\d$.

\begin{lemma}\label{lm3.6a}
The poles of the operators  $T_{8}^\pm$ coincide with the
eigenvalues of the operator $H(a)$. For any compact set
$K\subset \mathbb{S}_\d$ separated from $\discspec(H(a))$ by a
positive distance the estimates
\begin{equation}\label{3.21a}
\left\|\frac{\p^i T_{8}^\pm}{\p\l^i}\right\|\leqslant Cl^i\E^{-2
l \RE \kappa_1^+(\l)},\quad i=0,1,
\end{equation}
hold true with $C$ which is independent of $\l\in K$ and $l$.  For
any $\l$ close to an eigenvalue $\l_n$ of the operator $H(a)$ the
representation
\begin{equation}
T_{8}^\pm(\l,l,a,\widetilde{a})=\frac{\phi_n^\pm}{\l-\l_n}
T_{4}(a) + T_{9}^\pm(\l,l,a,\widetilde{a})\label{3.17}
\end{equation}
is valid, where
\begin{equation}\label{3.17a}
\phi_n^\pm(x_1,l,a):=\frac{\p\psi_n}{\p x_2}(x_1\pm
2l,+0,a)-\frac{\p\psi_n}{\p x_2}(x_1\pm 2l,-0,a),\quad
x_1\in(-\widetilde{a},\widetilde{a}).
\end{equation}
The operators $T_{9}^\pm: L_2(\g_a)\to L_2(\g_{\widetilde{a}})$
are bounded and holomorphic w.r.t. $\l$ in the vicinity of
$\l_n$ and satisfy the estimates
\begin{equation}\label{3.16}
\left\|\frac{\p^i T_{9}^\pm}{\p\l^i}\right\|\leqslant
Cl^{i+1}\E^{-2l\RE \kappa_1^+(\l)},\quad i=0,1,
\end{equation}
where the constant $C$ is independent of $\l$ and $l$.
\end{lemma}
\begin{proof}
Due to Lemma~\ref{lm3.12} we have
\begin{equation}\label{3.27a}
T_{8}^\pm(\l,l,a,\widetilde{a})f=T_{7}^\pm(\l,l,a,\widetilde{a})u.
\end{equation}
Here $u$ is a solution to the boundary value problem
(\ref{3.11}). Using this identity and the representation
(\ref{3.9}), we arrive at (\ref{3.17}), where $T_{9}^\pm$ is
bounded operator holomorphic in $\lambda$. Moreover,
\begin{align*}
T_{9}^\pm(\l,l,a,\widetilde{a})=&\frac{
T_{7}^\pm(\l,l,a,\widetilde{a})-
T_{7}^\pm(\l_n,l,a,\widetilde{a})}{\l-\l_n}
T_{4}(a)+T_{7}^\pm(\l,l,a,\widetilde{a})T_{5}^\pm(\l,a)
\\
=&\left(\frac{1}{\l-\l_n}\int\limits_{\l_n}^{\l}\frac{\p
T_{7}^\pm}{\p\l}(z,l,a,\widetilde{a})\di z\right)
T_{4}(a)+T_{7}^\pm(\l,l,a,\widetilde{a})T_{5}^\pm(\l,a),
\\
\frac{T_{9}^\pm}{\p\l}(\l,l,a,\widetilde{a})=&
\left(\frac{1}{(\l-\l_n)^2}\int\limits_{\l_n}^{\l}
\int\limits_{z_1}^{\l}\frac{\p^2
T_{7}^\pm}{\p\l^2}(z_2,l,a,\widetilde{a})\di z_2\di z_1\right)
T_{4}(a)
\\
&\hphantom{\Bigg(}+\frac{\p}{\p\l}\left(
T_{7}^\pm(\l,l,a,\widetilde{a})T_{5}^\pm(\l,a)\right).
\end{align*}
Applying now Lemma~\ref{lm3.7} we obtain the estimates
(\ref{3.16}).

The operators $\frac{\p^i T_{3}}{\p\l^i}$, $i=0,1$, are bounded
uniformly in $\l\in K$, thus in view of the relation
(\ref{3.27a}) and Lemma~\ref{lm3.7} we arrive readily at the
estimates (\ref{3.21a}).
\end{proof}

Concluding this section we shall prove Propositions~\ref{lm1.2}
and \ref{lm1.3}.
\begin{proof}[Proof of Proposition~\ref{lm1.2}]
Applying Lemma~\ref{lm3.12} to $\psi_n$ with $b_1=\pm a$,
$b_2=0$, $b_3=\pi$ and $b_2=-d$, $b_3=0$, we obtain the
formul{\ae} (\ref{1.4}), (\ref{1.4a}). The factor $(\pm
1)^{n+1}$ in these formul{\ae} is due to the definite parity of
$\psi_n$ w.r.t. $x_1$. The formula (\ref{1.5}) for $c_1^{(n)}$
follows from the chain of relations obtained by integration by
parts,
\begin{align*}
0&=\int\limits_{\Pi^+}\E^{\pm \kappa_1^+(\l_n)x_1}\sin
x_2(\D+\l_n)\psi_n(x,a)\di x
\\
&=\int\limits_{\g_a}\E^{\pm \kappa_1^+(\l_n)x_1}\psi_n(x,a)\di
x-(\pm 1)^{n+1}\pi \kappa_1^+(\l_n)c(\l_n,a).
\end{align*}
It remains to check the inequalities $c(\l_i,a)\not=0$, $i=1,2$.
The eigenfunction $\psi_1$ associated with the ground state can be
chosen non-negative. Moreover, $\psi_1$ is not identically zero at
$\g_a$, since otherwise it would be an eigenfunction of the
negative Dirichlet Laplacian in $\Pi^+$ and would correspond to
the eigenvalue $\l_1<1$. At the same time, the spectrum of the
mentioned operator is the halfline $[1,+\infty)$. The described
properties of $\psi_1$ and the formula (\ref{1.15}) imply that
$c(\l_1,a)\not=0$.

According to Proposition~\ref{lm1.1} the eigenfunction $\psi_2$
is odd w.r.t. $x_1$. It allows us to modify the formula
(\ref{1.5}),
\begin{equation}\label{3.20}
\begin{aligned}
c(\l_2,a)=&\frac{1}{\pi \kappa_1^+(\l_2)}\int\limits_{\g_a}
\psi_2(x,a)\sinh \kappa_1^+(\l_2)x_1\di x_1
\\
=&\frac{2}{\pi \kappa_1^+(\l_2)}
\int\limits_0^a\psi_2(x_1,0,a)\sinh \kappa_1^+(\l_2)x_1\di x_1.
\end{aligned}
\end{equation}
The  eigenvalue $\l_2$ is the ground state of the negative
Laplacian in $\Pi_a\cap\{x: x_1>0\}$ subject to Dirichlet
boundary condition on $\p\Pi_a\cap\{x: x_1>0\}$ and to Neumann
one on $\{x: x_1=0, -d<x_2<\pi\}$. Hence
$\psi_2(x_1,0,a)\geqslant 0$, and $\psi_2(x_1,0,a)\not\equiv0$
as $x_1\in(0,a)$, and by (\ref{3.20}) these inequalities imply
that $c(\l_2,a)\not=0$.
\end{proof}
\begin{proof}[Proof of Proposition~\ref{lm1.3}]
The unique solvability of the problem (\ref{1.3a}) is ensured by
Lemma~\ref{lm3.13}. Moreover, we have
$U=T_{3}(\l,a)\E^{\kappa_1^+(\l)x_1}$. The relations
(\ref{1.3b}), (\ref{1.3c}) follow from Lemma~\ref{lm3.12}, and
the formula (\ref{1.10}) is proved in the same way as
(\ref{1.5}).

Integrating by parts and employing the formula (\ref{1.10}), we
obtain a chain or identities,
\begin{align*}
0=&\int\limits_\Pi U(\D+\l)U\di x=\l\|U\|_{L_2(\Pi)}^2-\|\nabla
U \|_{L_2(\Pi)}^2-
\\
&-\int\limits_{\g_a} U\left(\frac{\p U}{\p x_2}\Big|_{x_2=+0}-
\frac{\p U}{\p x_2}\Big|_{x_2=-0}\right)\di x_1
\\
=&\l\|U\|_{L_2(\Pi)}^2-\|\nabla U \|_{L_2(\Pi)}^2-\pi
\kappa_1^+(\l) c(\l,a),
\end{align*}
which implies
\begin{equation}\label{3.21}
c(\l,a)=\frac{\l\|U\|_{L_2(\Pi)}^2-\|\nabla U
\|_{L_2(\Pi)}^2}{\pi \kappa_1^+(\l)}.
\end{equation}
Since $U\in\H^1(\Pi,\p\Pi)$, the minimax principle yields the
inequality
\begin{equation*}
\|U\|_{L_2(\Pi)}^2\leqslant \frac{1}{\l_1(a)}\|\nabla U
\|_{L_2(\Pi)}^2.
\end{equation*}
We substitute this inequality into the formula (\ref{3.21}) and
obtain
\begin{equation*}
c(\l,a)\leqslant \frac{1}{\pi
\kappa_1^+(\l)}\left(\frac{\l}{\l_1(a)}-1\right)\|\nabla U
\|_{L_2(\Pi)}^2<0,
\end{equation*}
if $\l<\l_1(a)$.
\end{proof}

\section{Reduction of the perturbed problem} \label{s: twowin}

After this preliminary let us turn to our main problem; we are
going to reformulate it as a suitable operator equation. Recall
that we are looking for eigenvalues of the operator $H$, i.e.
non-trivial $L_2(\Pi)$-solutions to the boundary value problem
\begin{equation}\label{4.1}
-\D\psi=\l\psi,\quad x\in\Pi,\qquad \psi=0,\quad x\in\p\Pi.
\end{equation}
We denote $Q^b:=\{x: -b<x_1<b, -d<x_2<\pi\}$ and introduce the
cut-off regions $\Pi^b:=\Pi\cap Q^b$, $\G^b:=\p\Pi\cap Q^b$.
Solutions to the problem (\ref{4.1}) can be identified with
functions belonging to $\H^1(\Pi^b,\G^b)$ for any $b>0$ such
that
\begin{equation}\label{4.0}
(\nabla\psi,\nabla\z)_{L_2(\Pi)}=\l(\psi,\z)_{L_2(\Pi)},
\end{equation}
holds for each $\z\in C^\infty_0(\Pi)$; it follows from the
smoothness-improving theorem mentioned above that such a $\psi$
belongs to $C^\infty(\Pi)$.

We assume that $\l\in \mathbb{S}_\d$, with $\d>0$ is chosen in
such a way that $\sigma_*\subset \mathbb{S}_\d$. Let
$f_\pm=f_\pm(\cdot,l)\in L_2(\g_{a_\pm})$ be an arbitrary pair of
functions. Denote by $u_\pm$ the solutions of the problem
(\ref{3.11}) with $a=a_\pm$ and $f=f_\pm\in L_2(\g_{a_\pm})$ and
assume that $u_\pm\in L_2(\Pi_{a_\pm})$. We will seek a solution
to the problem (\ref{4.1}) in the form
\begin{equation}\label{4.2}
\psi(x,\l,l)=u_+(x_1-l,x_2,\l,l)+u_-(x_1+l,x_2,\l,l).
\end{equation}
Suppose for a moment that the function $\psi$ defined in this way
solves the problem (\ref{4.1}). In such a case the function $\psi$
is infinitely differentiable at the points of the segments
$\g_{a_\pm}$, and therefore
\begin{equation*}
\frac{\p\psi}{\p x_2}(x_1,+0,\l,l)-\frac{\p\psi}{\p
x_2}(x_1,-0,\l,l)=0,\quad x\in\g_\pm.
\end{equation*}
Substituting from (\ref{4.2}) into this identity, we obtain a pair
of equations,
\begin{equation}\label{3.4a}
f_\pm(x_1) + \frac{\p u_\mp}{\p x_2}(x_1\pm 2l,+0,\l,l)-\frac{\p
u_\mp}{\p x_2}(x_1\pm 2l,-0,\l,l)=0,\quad x\in\g_{a_\pm}.
\end{equation}
Denote $\boldsymbol{f}=(f_+,f_-)\in L_2(\g_{a_+})\oplus
L_2(\g_{a_-})$. The following lemma states that the last equation
is equivalent to the original problem (\ref{4.1}).

\begin{lemma}\label{lm4.1}
To any solution $\boldsymbol{f}\in L_2(\g_{a_+})\oplus
L_2(\g_{a_-})$ of (\ref{3.4a}) and functions $u_\pm$ solving
(\ref{3.11}) for $a=a_\pm$, $f=f_\pm$ there exists a unique
$L_2(\Pi)$-solution of (\ref{4.1}) given by (\ref{4.2}).
Reversely, for any solution $\psi$ of (\ref{4.1}) there are
unique $\boldsymbol{f}\in L_2(\g_{a_+})\oplus L_2(\g_{a_-})$
solving (\ref{3.4a}) and unique functions $u_\pm\in
L_2(\Pi_{a_{\pm}})$ satisfying (\ref{3.11}) with $a=a_\pm$,
$f=f_\pm$ such that $\psi$ is given by (\ref{4.2}). This
equivalence holds for any $\l\in \mathbb{S}_\d$ and
$l\geqslant\max\{a_-,a_+\}+1$.
\end{lemma}
\begin{proof}
Suppose that $\boldsymbol{f}\in L_2(\g_{a_+})\oplus
L_2(\g_{a_-})$ is a solution to the equations (\ref{3.4a}),
where the functions $u_\pm\in L_2(\Pi_{a_\pm})$ solve the
problem (\ref{3.11}) $a=a_\pm$ and $f=f_\pm$. We define $\psi$
in accordance with (\ref{4.2}). The functions $u_\pm$ are
elements of $L_2(\Pi)$, hence the same is true for $\psi$.
Moreover, the function $\psi$ belongs obviously to
$\H^1(\Pi^b,\G^b)$ for each $b>0$ and vanishes on $\G$.

Let us check that the function $\psi$ satisfies the equation
(\ref{4.0}). To this purpose, we indicate by
$\chi_2=\chi_2(x_1)$ an infinitely differentiable cut-off
function being equal to one if $|x_1+l|<\max\{a_+,a_-\}+1/2$ and
vanishing if $|x_1+l|>\max\{a_+,a_-\}+1$. For any $\z\in
C_0^\infty(\Pi)$ we have
\begin{equation}\label{4.4}
\begin{aligned}
&\big(\nabla
u_+(x_1-l,x_2,\l,l),\nabla\z\big)_{L_2(\Pi)}-\l\big(u_+(x_1-l,x_2,\l,l),
\z\big)_{L_2(\Pi)}
\\
&=\big(\nabla
u_+(x_1-l,x_2,\l,l),\nabla(\z\chi_2)\big)_{L_2(\Pi)}-
\l\big(u_+(x_1-l,x_2,\l,l), \z\chi_2)\big)_{L_2(\Pi)}
\\
&+\big(\nabla
u_+(x_1-l,x_2,\l,l),\nabla(\z(1-\chi_2))\big)_{L_2(\Pi)}-
\l\big(u_+(x_1-l,x_2,\l,l), \z(1-\chi_2)\big)_{L_2(\Pi)}.
\end{aligned}
\end{equation}
Since $u_+(\cdot)$ is an element of $C^\infty(\Pi_{a_+}\setminus
\overline{Q}^{a_+})$, we can integrate by parts,
\begin{align*}
&\big(\nabla
u_+(x_1-l,x_2,\l,l),\nabla\z\chi_2)\big)_{L_2(\Pi)}-
\l\big(u_+(x_1-l,x_2,\l,l), \z\chi_2\big)_{L_2(\Pi)}
\\
&=-\int\limits_{\g_-}\left(\frac{\p u_+}{\p
x_2}(x_1-l,+0,\l,l)-\frac{\p u_+}{\p
x_2}(x_1-l,-0,\l,l)\right)\z\di
x_1-
\\
&-\int\limits_\Pi\z\chi_2(\D+\l) u_+(x_1-l,x_2,\l,l)\di x=
\int\limits_{\g_-} f_-(x_1+l)\z\di x_1.
\end{align*}
We have employed here the equation satisfied by $u_+$ as well as
the equation (\ref{3.4a}) for $f_+$. Since
$\z(x_1+l,x_2)(1-\chi_2(x_1+l))\in C^\infty_0(\Pi_{a_+})$, we
can use the identity (\ref{3.12}) to infer that
\begin{align*}
\big(\nabla
u_+(x_1-l,x_2,\l,l),\nabla\z(1-\chi_2)\big)_{L_2(\Pi)}-&
\big(u_+(x_1-l,x_2,\l,l),\z(1-\chi_2)\big)_{L_2(\Pi)}
\\
&=-\int\limits_{\g_+}f_+(x_1-l)\z\di x_1.
\end{align*}
We substitute now the last two relations into (\ref{4.4}) and
arrive at the identity
\begin{align*}
&\big(\nabla u_+(x_1-l,x_2,\l,l),\nabla\z\big)_{L_2(\Pi)}-
\l\big(u_+(x_1-l,x_2,\l,l), \z\big)_{L_2(\Pi)}
\\
&=(f_-(x_1+l),\z)_{L_2(\g_-)}-(f_+(x_1-l),\z)_{L_2(\g_+)}.
\end{align*}
In the same way one can check that
\begin{align*}
&\big(\nabla u_-(x_1+l,x_2,\l,l),\nabla\z\big)_{L_2(\Pi)}-
\l\big(u_-(x_1+l,x_2,\l,l), \z\big)_{L_2(\Pi)}
\\
&=(f_+(x_1-l),\z)_{L_2(\g_+)}-(f_-(x_1+l),\z)_{L_2(\g_-)};
\end{align*}
summing the last two relations we arrive at the relation
(\ref{4.0}) for the function $\psi$.

Let $\psi$ be a solution to the problem (\ref{4.1}) belonging to
$L_2(\Pi)$. By smoothness-improving theorems the function $\psi$
belongs to $C^\infty(\{x: -1\leqslant x_1\leqslant 1, 0\leqslant
x_2\leqslant \pi\})$ and to $C^\infty(\{x: -1\leqslant
x_1\leqslant 1, -d\leqslant x_2\leqslant 0\})$. This allows us to
define the numbers
\begin{align*}
&\a_j^\pm=\a_j^\pm(\l,l):=
\frac{2}{\pi}\int\limits_0^\pi\left(\psi(0,x_2,\l,l)\pm
\frac{1}{\kappa_j^+}\frac{\p\psi}{\p x_1}(0,x_2,\l,l)\right)\sin
j x_2\di x_2,
\\
&\b_j^\pm=\b_j^\pm(\l,l):=
\frac{2}{d}\int\limits_0^\pi\left(\psi(0,x_2,\l,l)\pm
\frac{1}{\kappa_j^-}\frac{\p\psi}{\p x_1}(0,x_2,\l,l)\right)\sin
\frac{\pi j}{d}x_2\di x_2.
\end{align*}
Using these numbers, we introduce the functions $u_\pm$ in the
following way:
\begin{align*}
&u_\pm(x_1\mp l,x_2,\l,l):=\sum\limits_{j=1}^{\infty}
\a_j^\pm(\l,l)\E^{\pm \kappa_j^+ x_1}\sin j x_2,&& \pm
x_1\leqslant 0,\quad x_2\in(0,\pi),
\\
&u_\pm(x_1\mp l,x_2,\l,l):=\sum\limits_{j=1}^{\infty}
\b_j^+(\l,l)\E^{\pm \kappa_j^- x_1}\sin \frac{\pi j}{d} x_2,&&
\pm x_1\leqslant 0,\quad x_2\in(-d,0),
\\
&u_\pm(x_1\mp l,x_2,\l,l):=\psi(x,\l,l)-u_\pm(x_1\pm l,x_2,\l),&&
\pm x_1>0,\quad x_2\in(-d,\pi).
\end{align*}
Proceeding in the same way as in the proof of Lemma~4.1 in
\cite{B2}, we check that the functions $u_\pm$ are well defined
and
\begin{align}\label{4.6}
&u_\pm\in \H^1(\Pi_{a_\pm},\G_{a_\pm})\cap\H^2(S), \quad
S\in\Xi_a,
\\
&(\D+\l)u_\pm(x_1\mp l,x_2,\l,l)=0,\quad x\in\Pi\setminus\{x:
x_1=0\}.\label{4.8}
\end{align}
The relation (\ref{4.2}) follows from the definition of the
functions $u_\pm$. Now we set
\begin{equation}\label{4.8a}
f_\pm(x_1,l):=-\frac{\p u_\mp}{\p x_2}(x_1\pm 2l,+0,\l,l)+
\frac{\p u_\mp}{\p x_2}(x_1\pm 2l,-0,\l,l),\quad
x_1\in(-a_\pm,a_\pm);
\end{equation}
in view of (\ref{4.6}) we can conclude that $f_\pm\in
L_2(\g_{a_\pm})$. We also note that the definition of  $u_\pm$
and the smoothness of $\psi$ at $\g_\pm$ imply
\begin{equation}\label{4.7}
f_\pm(x_1,l)=\frac{\p u_\pm}{\p x_2}(x_1,+0,\l,l)-\frac{\p
u_\pm}{\p x_2}(x_1,-0,\l,l),\quad x_1\in(-a_\pm,a_\pm).
\end{equation}

Let us check the integral equation (\ref{3.12}) for the function
$u_+=u_+(x,\l)$. Taking into account  (\ref{4.8}), (\ref{4.7}),
and integrating by parts, we get
\begin{align*}
&-(\nabla u_+,\nabla\z)_{L_2(\Pi_a)}+\l(u_+,\z)_{L_2(\Pi_a)}
\\
&= \left(\frac{\p u_+}{\p x_2}(x_1,+0,\l)- \frac{\p u_+}{\p
x_2}(x_1,-0,\l),\z\right)_{L_2(\g_{a_+})}=(f_+,\z)_{L_2(\g_{a_+})}.
\end{align*}
for any $\z\in C_0^\infty(\Pi_{a_+})$. In the same way one can
check that
\begin{equation*}
-(\nabla u_-,\nabla\z)_{L_2(\Pi_a)}+\l(u_-,\z)_{L_2(\Pi_a)}
=(f_-,\z)_{L_2(\g_{a_-})}
\end{equation*}
for any $\z\in C_0^\infty(\Pi_{a_-})$, thus $u_\pm$ are
solutions to the problem (\ref{3.11}) with $a=a_\pm$ and
$f=f_\pm$. The equations (\ref{3.4a}) follow from (\ref{4.8a}).
\end{proof}

Suppose that $\l\in \mathbb{S}_\d\setminus\sigma_*$. In that
case the functions $u_\pm$ introduced above can be represented
as $u_\pm=T_{3}(\l,a_\pm)f_\pm$, thus the equations (\ref{3.4a})
become
\begin{equation}\label{4.3a}
\boldsymbol{f}+T_{8}(\l,l,\boldsymbol{a})\boldsymbol{f}=0,
\end{equation}
where the operator $T_{8}: L_2(\g_{a_+})\oplus L_2(\g_{a_-})\to
L_2(\g_{a_+})\oplus L_2(\g_{a_-})$ is defined by
\begin{equation*}
T_{8}(\l,l,\boldsymbol{a})\boldsymbol{f}:=\big(
T_{8}^+(\l,l,a_-,a_+)f_-,T_{8}^-(\l,l,a_+,a_-)f_+\big).
\end{equation*}
Now we are ready to demonstrate the first one of our main
results.

\begin{proof}[Proof of Theorem~\ref{th1.0}]
If $a_\pm=0$ the essential spectrum of the operator $H$ is
obviously $[1,+\infty)$, and an elementary argument using
Dirichlet-Neumann bracketing \cite[Ch. X\!I\!I\!I, \S 15]{RS} and
the minimax principle \cite[Ch. X\!I\!I\!I, \S 1]{RS} shows that
the threshold of the essential spectrum of $H$ is one, i.e.
$\essspec(H)\subseteq[1,+\infty)$. The opposite inclusion can be
shown easily; one needs to employ Weyl's criterion (see, for
instance, proof of Lemma~2.1 in \cite{B2}).

The operator $H$ being self-adjoint, its isolated eigenvalues are
real, and in view of the above observation they are smaller than
one; we arrange them conventionally in the ascending order
counting multiplicity. Next we use bracketing again in a way
analogous to \cite{ESTV}: we add Neumann boundaries at segments
corresponding to $x_1$ at the endpoints of $\g_\pm$ and $x_2\in
(-d,\pi)$. In this way we get an operator estimating $H$ from
below, and since only the window parts contribute to the spectrum
below one we infer by minimax that $H$ has finitely many
eigenvalues for any $l>0$ and their number has a bound independent
of $l$.

Let $K\subset \mathbb{S}_\d$ be any compact set separated from
$\sigma_*$ by a positive distance. By the estimates
(\ref{3.21a}) the operator $T_{8}$ has a norm being strictly
less than one for $\l\in K$ and $l$ large enough. For such $\l$
and $l$ the equation (\ref{4.3a}) has thus a trivial solution
only, and in view of Lemma~\ref{lm4.1} this implies that the
operator $H$ has no eigenvalues in the set $K$ if $l$ is large
enough. This means that each eigenvalue of the operator $H$ has
to converge to one of the numbers from the set $\sigma_*$ or to
the threshold of the essential spectrum.
\end{proof}

The eigenvalues $H$, i.e. those $\l$ for which the problem
(\ref{4.1}) has a nontrivial $L_2(\Pi)$-solution, coincide in view
of Lemma~\ref{lm4.1} with the values of $\l$ for which the
equation (\ref{3.4a}) has a nontrivial solution. In the case
considered here we deal only with the eigenvalues of $H$ which
converge to a value $\l_*\in\sigma_*$ separated from the
threshold, in other words, being smaller than one.

Our aim is to solve the equation (\ref{3.4a}) and to obtain in
this way an equation for the aforementioned values of $\l$.
Consider a $\l_*\in\sigma_*$; if $\l_*=\l_n$ is an eigenvalue of
the operator $H(a_+)$ we set
\begin{equation*}
\boldsymbol{\phi}_*^+(\cdot,l):=\big(0,\phi_n^-(\cdot,l,a_+)\big)\in
L_2(\g_{a_+})\oplus L_2(\g_{a_-}),\quad
T_{4}^+\boldsymbol{f}:=(f_+,\psi_n)_{L_2(\g_{a_+})},
\end{equation*}
where $\phi_n^-$ is determined by $\psi_n$ in accordance with
(\ref{3.17a}) and $\psi_n$ is an eigenfunction associated with
$\l_n$, in the opposite case we put
\begin{equation*}
\boldsymbol{\phi}_*^+(\cdot,l):=(0,0)\in L_2(\g_{a_+})\oplus
L_2(\g_{a_-}),\quad T_{4}^+\boldsymbol{f}:=0.
\end{equation*}
Analogously, if $\l_*=\l_n$ is an eigenvalue of $H(a_-)$ we set
\begin{equation*}
\boldsymbol{\phi}_*^-(\cdot,l):=\big(\phi_n^+(\cdot,l,a_-),0\big)\in
L_2(\g_{a_+})\oplus L_2(\g_{a_-}),\quad
T_{4}^-\boldsymbol{f}:=(f_-,\psi_n)_{L_2(\g_{a_-})},
\end{equation*}
where $\phi_n^+$ corresponds to $\psi_n$ according to
(\ref{3.17a}) and $\psi_n$ is an eigenfunction associated with
$\l_n$, otherwise
\begin{equation*}
\boldsymbol{\phi}_*^-(\cdot,l):=(0,0)\in L_2(\g_{a_+})\oplus
L_2(\g_{a_-}),\quad T_{4}^-\boldsymbol{f}:=0.
\end{equation*}

Given a number $\l_*\in\si_*$, we consider the equation
(\ref{3.4a}) for $\l$ in the vicinity of $\l_*$. Assume first
that $\l\ne\l_*$, in which case the equation (\ref{3.4a}) is
equivalent to (\ref{4.3a}). In view of Lemma~\ref{lm3.6a} the
operator $T_{8}$ is bounded and meromorphic as a function of
$\l\in \mathbb{S}_\d$, and the numbers $\l_*\in\sigma_*$ are
poles of $T_{8}$. For any $\l$ close to $\l_*$ the operator
$T_{8}$ can be thus represented as
\begin{equation}\label{4.12}
T_{8}(\l,l,\boldsymbol{a})
=\boldsymbol{\phi}_*^+(\cdot,l)\frac{T_{4}^+ }{\l-\l_*} +
\boldsymbol{\phi}_*^-(\cdot,l)\frac{T_{4}^- }{\l-\l_*}
+T_{9}(\l,l,\boldsymbol{a}),
\end{equation}
where the operator $T_{9}$ acts as
\begin{equation*}
T_{9}(\l,l,\boldsymbol{a})\boldsymbol{f}:=\big(
T_{8}^+(\l,l,a_-,a_+)f_-,T_{9}^-(\l,l,a_+,a_-)f_+\big)
\end{equation*}
if $\l_*\in\discspec(H(a_+))\setminus\discspec(H(a_-))$,
\begin{equation*}
T_{9}(\l,l,\boldsymbol{a})\boldsymbol{f}:=\big(
T_{9}^+(\l,l,a_-,a_+)f_-,T_{8}^-(\l,l,a_+,a_-)f_+\big)
\end{equation*}
if $\l_*\in\discspec(H(a_-))\setminus\discspec(H(a_+))$, and
finally,
\begin{equation*}
T_{9}(\l,l,\boldsymbol{a})\boldsymbol{f}:=\big(
T_{9}^+(\l,l,a_-,a_+)f_-,T_{9}^-(\l,l,a_+,a_-)f_+\big)
\end{equation*}
if $\l_*\in\discspec(H(a_-))\cap\discspec(H(a_+))$. The operator
$T_{9}$ on $L_2(\g_{a_+})\oplus L_2(\g_{a_-})$ is bounded and
holomorphic w.r.t. $\l$ in the vicinity of $\l_*$, and the
estimate
\begin{equation}\label{4.13}
\left\|\frac{\p^i T_{9}}{\p\l^i}\right\|\leqslant
Cl^{i+1}\E^{-2l\RE \kappa_1^+(\l)},\quad i=0,1,
\end{equation}
holds true with a constant $C$ which is independent on $\l$ and
$l$.

We substitute the representation (\ref{4.12}) into (\ref{4.3a})
to obtain
\begin{equation*}
\boldsymbol{f}+\frac{T_{4}^+\boldsymbol{f}}{\l-\l_*}
\boldsymbol{\phi}_*^++ \frac{T_{4}^-\boldsymbol{f}}{\l-\l_*}
\boldsymbol{\phi}_*^-+T_{9}\boldsymbol{f}=0.
\end{equation*}
Since the norm of $T_{9}$ is small for large $l$ due to
(\ref{4.13}), the operator $(\I+T_{9})^{-1}$ is well defined
being bounded in $L_2(\g_{a_+})\oplus L_2(\g_{a_-})$. We apply
this operator to the last equation arriving at
\begin{equation}\label{4.15}
\boldsymbol{f}+\frac{T_{4}^+\boldsymbol{f}}{\l-\l_*}\boldsymbol{\Phi}_*^+
+ \frac{T_{4}^-\boldsymbol{f}}{\l-\l_*}\boldsymbol{\Phi}_*^-=0,
\end{equation}
where $\boldsymbol{\Phi}_*^\pm(\cdot,\l,l)= (\I+T_{9}(\l,l,
\boldsymbol{a}))^{-1} \boldsymbol{\phi}_*^\pm(\cdot,l)$. The
last equation implies that
\begin{equation}\label{4.17a}
\boldsymbol{f}=c_+\boldsymbol{\Phi}^+_*+ c_-\boldsymbol{\Phi}^-_*
\end{equation}
for some  numbers $c_\pm$. We substitute from here into
(\ref{4.15}) obtaining
\begin{equation}\label{4.18}
\boldsymbol{\Phi}_*^+\left(c_+\left(1+\frac{A_{11}}{\l-\l_*}\right)
+c_-\frac{A_{12}}{\l-\l_*}\right)+\boldsymbol{\Phi}_*^-
\left(c_+\frac{A_{21}}{\l-\l_*}+
c_-\left(1+\frac{A_{22}}{\l-\l_*}\right)\right)=0,
\end{equation}
where the quantities $A_{ij}=A_{ij}(\l,l)$ are defined by
\begin{gather*}
A_{11}(\l,l):=T_{4}^+\Phi_*^+(\cdot,\l,l),\quad
A_{12}(\l,l):=T_{4}^+\Phi_*^-(\cdot,\l,l),
\\
A_{21}(\l,l):=T_{4}^-\Phi_*^+(\cdot,\l,l),\quad
A_{22}(\l,l):=T_{4}^-\Phi_*^-(\cdot,\l,l).
\end{gather*}
The definition of $\boldsymbol{\Phi}_*^\pm$ together with the
estimate (\ref{4.13}) imply for $l$ large enough
\begin{equation}\label{4.16a}
\boldsymbol{\Phi}_*^\pm=\boldsymbol{\phi}_*^\pm+
\Odr\big(\E^{-2l\RE
\kappa_1^+(\l)}\|\boldsymbol{\phi}_*^\pm\|\big).
\end{equation}
If $\boldsymbol{\phi}_*^+\not=0$, and $\boldsymbol{\phi}_*^-=0$,
in particular, we have
\begin{equation}\label{4.19}
\boldsymbol{\Phi}_*^+\not=0,\quad  \boldsymbol{\Phi}_*^-=0,\quad
A_{12}=A_{22}=0,
\end{equation}
and in this case the equation (\ref{4.18}) holds if and only if
\begin{equation*}
c_+\left(1+\frac{A_{11}}{\l-\l_*}\right)=0.
\end{equation*}
If $\boldsymbol{f}$ corresponds to an eigenfunction $\psi$ of
the problem (\ref{4.1}) by (\ref{4.2}), the number $c_+$ is
non-zero. Indeed, in the opposite case (\ref{4.17a}) and
(\ref{4.19}) would imply that $\boldsymbol{f}=0$, which by
Lemma~\ref{lm4.1} results in $\psi=0$. Consequently, the
equation (\ref{4.3a}) has in this case a nontrivial solution if
and only if
\begin{equation}\label{4.20}
\l-\l_*+A_{11}(\l,l)=0.
\end{equation}
If $\l$ is a root of this equation, the corresponding nontrivial
solution of (\ref{4.3a}) can be expressed as (\ref{4.17a}) with
$c_+\not=0$ and $c_-=0$.

In the case $\boldsymbol{\phi}_*^+=0$ and
$\boldsymbol{\phi}_*^-\not=0$ similar arguments lead us to the
conclusion that the equation (\ref{4.3a}) has a nontrivial
solution if and only if
\begin{equation}\label{4.21}
\l-\l_*+A_{22}(\l,l)=0,
\end{equation}
and the corresponding non-trivial solution can be written as
(\ref{4.17a}) with the coefficients $c_+=0$ and $c_-\not=0$.

Finally, if both the functions $\boldsymbol{\phi}_*^\pm$ are
non-zero, they are linearly independent by definition and the same
is true for the functions $\boldsymbol{\Phi}_*^\pm$. Hence the
equation (\ref{4.3a}) holds if and only if
\begin{equation}\label{4.16}
\left((\l-\l_*)\mathrm{E}+
\mathrm{A}(\l,l)\right)\boldsymbol{c}=0,
\end{equation}
where $\mathrm{E}$ is the unit matrix, and
\begin{equation*}
\mathrm{A}(\l,l):=
\begin{pmatrix}
A_{11}(\l,l)& A_{12}(\l,l)
\\
A_{21}(\l,l)& A_{22}(\l,l)
\end{pmatrix},
\quad \boldsymbol{c}:=
\begin{pmatrix}
c_+
\\
c_-
\end{pmatrix}.
\end{equation*}
The column $\boldsymbol{c}$ is non-zero, since otherwise
(\ref{4.17a}) and (\ref{4.19}) would imply $\boldsymbol{f}=0$,
thus the system (\ref{4.16}) of linear equations has a nontrivial
solution if and only if
\begin{equation}\label{4.20a}
\det \left((\l-\l_*)\mathrm{E}+\mathrm{A}(\l,l)\right)=0,
\end{equation}
which can be rewritten as
\begin{equation}\label{4.17}
(\l-\l_*)^2+(\l-\l_*)\tr \mathrm{A}(\l,l)+\det
\mathrm{A}(\l,l)=0\,;
\end{equation}
the corresponding non-trivial solution of the equation
(\ref{4.3a}) is given by (\ref{4.17a}), where
$\left(\begin{smallmatrix} c_+ \\ c_-
\end{smallmatrix}\right)$ is a nontrivial solution of
(\ref{4.16}).

Assume now that $\l=\l_*$. Let $\l_*$ coincide with an
eigenvalue $\l_n$ of the operator $H(a_+)$ being not at the same
time an eigenvalue of $H(a_-)$. In this case we again can claim
that $u_-=T_{3}(\l_*,a_-)f_-$, on the other hand, the boundary
value problem for $u_+$ with $\l=\l_*$ is solvable if and only
if
\begin{equation}\label{3.50}
0=\int\limits_{\g_{a_+}} f_+\psi_n(x,a_+)\di
x=T_{4}^+\boldsymbol{f}.
\end{equation}
This follows from Lemma~\ref{lm3.13}. The function $u_+$ is
given by $u_+=T_{5}(\l_*,a_+)f_+ - c_+\psi_+$, where $c_+$ is a
constant. We can substitute now the described $u_\pm$ into
(\ref{3.4a}) and obtain
\begin{gather}
\boldsymbol{f}+T_{9}(\l,l,\boldsymbol{a})\boldsymbol{f}=
c_+\boldsymbol{\phi}_*^+,\nonumber
\\
\boldsymbol{f}=c_+\boldsymbol{\Phi}_*^+.\label{3.51}
\end{gather}
This function will generate a solution to the problem
(\ref{4.1}) if and only if (\ref{4.20}) holds true. Substituting
(\ref{3.51}) into (\ref{3.50}), we arrive at the equation
(\ref{4.20}) with $\l=\l_*$. If $c_+\not=0$ holds in
(\ref{3.51}) we see that the formula (\ref{3.50}) coincides with
(\ref{4.17a}) with $c_-=0$. Consequently, in the case
$\l_*\in\spec(H(a_-))\setminus\spec(H(a_+))$ the equation
(\ref{4.20}) determines all the values of $\l$ in the vicinity
of $\l_*$ for which the equation (\ref{3.4a}) has a nontrivial
solution; these nontrivial solutions are given by (\ref{4.17a})
with $c_+\not=0$ and $c_-=0$.

In the same way one can check that the equation (\ref{4.21})
determines the sought values of $\l$ in the case when $\l_*$ is an
eigenvalue of the operator $H(a_-)$ and not of $H(a_+)$. The
corresponding nontrivial solutions of (\ref{3.4a}) have $c_+=0$
and $c_-\not=0$.

Finally, if $\l_*\in\sigma_*$ is double and $\l=\l_*$, the
solvability conditions of the boundary value problems for $u$
are $T_{4}^\pm \boldsymbol{f}=0$. If this holds true, the
functions $u_\pm$ are given by $u_\pm=T_{5}(\l,a_\pm) -c_\pm
\psi_\pm(\cdot)$, where $c_\pm$ are constants and $\psi_\pm$ are
the eigenfunctions of $H(a_\pm)$ associated with $\l_*$. The
equation (\ref{3.4a}) becomes
\begin{equation*}
\boldsymbol{f}+T_{9}(\l_*,l,\boldsymbol{a})\boldsymbol{f}=
c_+\boldsymbol{\phi}_*^++ c_-\boldsymbol{\phi}_*^-,
\end{equation*}
which yields the relation (\ref{4.17a}). The solvability
conditions $T_{4}^\pm \boldsymbol{f}=0$ are nothing else than
the system of linear equations (\ref{4.16}). In this way
(\ref{4.17a}), (\ref{4.16}), and (\ref{4.20a}) describe the
sought values of $\l$ in the vicinity of $\l_*$ and the
corresponding nontrivial solutions of (\ref{3.4a}).

\section{Proofs of Theorems~\ref{th1.2}-\ref{th1.4}}
\label{s: remain_proof}

Now we are going to demonstrate the remaining part of our
claims.
\begin{proof}[Proof of Theorem~\ref{th1.2}]
We will give the proof for the case $\l_*=\l_n(a_-)$, the
argument for $\l_*=\l_n(a_+)$ is similar. In accordance with the
results of the previous section the eigenvalue $\l^-(a,l)$, if
it exists, it must be a root of the equation (\ref{4.21}). Let
us prove first that there is a unique root which converges to
$\l_*$ as $l\to+\infty$. Proposition~\ref{lm1.2} implies that
the relation
\begin{equation}\label{5.1}
\phi_n^+(x_1,l,a_-)=\tau
c(\l_*,a_-)\E^{-2\kappa_1^+(\l_*)l}\E^{-\kappa_1^+(\l_*)x_1}+
\Odr(\E^{-2 \rho(\l_*) l}),
\end{equation}
holds in the norm of $L_2(\g_{a_+})$, hence by the definition of
$\boldsymbol{\phi}_*^-$ we have
\begin{equation}\label{5.2}
\boldsymbol{\phi}_*^-(x_1,l)=\tau
c(\l_*,a_-)\E^{-2\kappa_1^+(\l_*)l}
(\E^{-\kappa_1^+(\l_*)x_1},0)+ \Odr(\E^{-2 \rho(\l_*) l}).
\end{equation}
This formula in combination with the estimate (\ref{4.13}) lead
to the relation
\begin{equation}\label{5.1a}
A_{22}(\l,l)=\Odr(\E^{-2\kappa_1^+(\l_*)l}).
\end{equation}
Since $T_{9}$ is holomorphic w.r.t. $\l$ and has a small norm
for large $l$, we infer that the left-hand side of the last
equation is holomorphic in $\l$. For a small $\d$ take the
circle of those $\l$ such that $|\l-\l_*|=\d$. In view of
(\ref{5.1a}) the function $A_{22}$ satisfies the estimate
$|A_{22}|<\d$ if $l$ is large enough and $|\l-\l_*|=\d$; by
Rouch\'{e} theorem it implies that the function $\l\mapsto
\l-\l_*+A_{22}(\l,l)$ has the same number of zeros in the disk
$\{\l: |\l-\l_*|<\d\}$ as the function $\l\mapsto \l-\l_*$ does.
The number $\d$ is arbitrary, so we can conclude that there is a
unique root of the equation (\ref{4.21}) converging to $\l_*$ as
$l\to+\infty$. As a consequence, there exists a unique
eigenvalue of the operator $H$ converging to $\l_*$ as
$l\to+\infty$; we will denote this eigenvalue as
$\l^-(l,\boldsymbol{a})$. The estimate (\ref{5.1a}) implies at
the same time that
\begin{equation}\label{5.1b}
\l^-(l,\boldsymbol{a})-\l_*=\Odr(\E^{-2\kappa_1^+(\l_*)l}).
\end{equation}

Let us derive the asymptotic expansion (\ref{1.11}) for the
eigenvalue $\l^-(l,\boldsymbol{a})$. In order to do it, we will
need to know the asymptotic behavior for $A_{22}$ in a way more
precise than (\ref{5.1a}). For the sake of brevity we will write
shortly $\l$ instead of $\l^-(l,\boldsymbol{a})$. The relations
(\ref{5.2}) together with the estimates (\ref{4.13}),
(\ref{5.1b}) imply that
\begin{equation}
\begin{aligned}
&A_{22}(\l,l)=T_{4}^-(\I+T_{9}(\l,l,\boldsymbol{a}))^{-1}
\boldsymbol{\phi}_*^-(\cdot,l)
\\
&=T_{4}^-\boldsymbol{\phi}_*^-(\cdot,l)- T_{4}^-
T_{9}(\l,l,\boldsymbol{a})\boldsymbol{\phi}_*^-(\cdot,l)+
\Odr(\|T_{9}\|^2\|\boldsymbol{\phi}_*^-\|)
\\
&=-T_{4}^-
T_{9}(\l_*,l,\boldsymbol{a})\boldsymbol{\phi}_*^-(\cdot,l)+
\Odr\left(|\l-\l_*|\left\|\frac{\p
T_{9}}{\p\l}\right\|\|\boldsymbol{\phi}_*^-\|
+\|T_{9}\|^2\|\boldsymbol{\phi}_*^-\|\right)
\\
&=-T_{4}^-T_{8}^-(\l,l, \boldsymbol{a})\phi_n^+(\cdot,l,a_-)+
\Odr\left(|\l-\l_*|
l^2\E^{-4\kappa_1^+(\l_*)l}+l\E^{-6\kappa_1^+(\l_*)l}\right).
\end{aligned}\label{5.4a}
\end{equation}
Taking into account the estimate (\ref{3.21a}) for $\|T_{8}^-\|$
and the relation (\ref{5.1}), we can proceed with the
calculations obtaining
\begin{equation}\label{5.3}
\begin{aligned}
A_{22}(\l,l)=&-\tau c(\l,a_-)\E^{-2\kappa_1^+(\l_*)l}\big(
T_{8}^-(\l_*,l,a_+,a_-)
\E^{-\kappa_1^+(\l_*)x_1},\psi_*\big)_{L_2(\g_{a_-})}
\\
&+\Odr\left(|\l-\l_*|l^2\E^{-4\kappa_1^+(\l_*)l}+\E^{-2l(
\kappa_1^+(\l_*)+\rho(\l_*))}\right),
\end{aligned}
\end{equation}
where we have denoted $\psi_*(x)=\psi_n(x,a_-)$. In view of the
relation (\ref{3.27a}) the function
$T_{8}^-(\l_*,l,a_+,a_-)\E^{\kappa_1^+(\l_*)x_1}$ coincides with
$T_{7}^-(\l,l,a_+,a_-)u$, where $u$ is the solution to the
problem (\ref{3.11}) with $a=a_+$, $\l=\l_*$, and
$f=\E^{-\kappa_1^+(\l_*)x_1}$. It is clear that
$u(x)=U(-x_1,x_2,\l_*,a_+)$, and in view of (\ref{1.5}),
(\ref{1.3b}), (\ref{1.3c}) we obtain
\begin{align*}
\big(T_{8}^-(\l_*,l,&a_+,a_-)\E^{-\kappa_1^+(\l_*)x_1},
\psi_*\big)_{L_2(\g_{a_-})}
\\
&=c(\l_*,a_+)\E^{-2\kappa_1^+(\l_*)l}
\big(\E^{\kappa_1^+(\l_*)x_1},
\psi_*\big)_{L_2(\g_{a_-})}+\Odr(\E^{-2\rho(\l_*)l})
\\
&=\pi c(\l_*,a_+)c(\l_*,a_-)\kappa_1^+(\l_*)
\E^{-2\kappa_1^+(\l_*)l} +\Odr(\E^{-2\rho(\l_*)l}).
\end{align*}
Substituting these identities into (\ref{5.3}), we finally
arrive at the following formula,
\begin{equation*}
A_{22}(\l,l)=\mu^-(l,\boldsymbol{a})\E^{-4\kappa_1^+(\l_*)l}
+\Odr\left(|\l-\l_*| l^2\E^{-4\kappa_1^+(\l_*)l}+\E^{-2l(
\kappa_1^+(\l_*)+\rho(\l_*))}\right),
\end{equation*}
where $\mu^-(l,\boldsymbol{a})$ is defined by (\ref{1.12}). It
allows us to rewrite the equation (\ref{4.21}) as
\begin{equation*}
(\l-\l_*)\big(1+\Odr(l^2\E^{-4\kappa_1^+(\l_*)l})\big)=
\mu^-(l,\boldsymbol{a})\E^{-4\kappa_1^+(\l_*)l}
+\Odr\left(\E^{-2l( \kappa_1^+(\l_*)+\rho(\l_*))}\right)\,;
\end{equation*}
expressing $(\l-\l_*)$ from here we get the asymptotic expansion
(\ref{1.11}) and the formula (\ref{1.12}).

Next we have to prove the asymptotic expansion for the
eigenfunction associated with $\l^-$. The nontrivial solution of
the equation (\ref{3.4a}) is given by (\ref{4.17a}) with $c_+=0$
and $c_-=1$, i.e. as $\boldsymbol{f}=\boldsymbol{\Phi}_*^-$. We
substitute it into the relation $u_+=T_{3}(\l^+,a_+)f_+$ and
take into account (\ref{5.2}), (\ref{4.16a}); this yields
\begin{equation*}
u_+=T_{3}(\l^+,a_+)f_+=\Odr(\|\boldsymbol{\phi}_*^-\|)=
\Odr(\E^{-2\kappa_1^+(\l_*)l}),
\end{equation*}
which holds true in $\H^1(\Pi_{a_+})$ and in $\H^2(S)$ for each
$S\in\Xi_{a_+}$. If $\l^-\not=\l_*$, we obtain similarly with
the help of Lemma~\ref{lm3.13}
\begin{equation}\label{5.9a}
u_-=T_{3}(\l^-,a_-)f_-=\frac{\psi_*}{\l^--\l_*}
T_{4}^-\boldsymbol{\Phi}_*^-+T_{5}^-(\l^-,a_-)f_-= \frac{
A_{22}(\l^-,l)\psi_*}{\l^--\l_*}+
\Odr(\|\boldsymbol{\phi}_*^-\|).
\end{equation}
Due to the equation (\ref{4.21}) it follows that
\begin{equation*}
u_-=-\psi_*+\Odr(\E^{-2\kappa_1^+(\l_*)l})
\end{equation*}
holds in $\H^1(\Pi_a)$ and $\H^2(S)$ for each $S\in\Xi_{a_-}$.
If $\l^-=\l_*$, the last relation holds again; in order to prove
it, it is sufficient to employ the identity
\begin{equation*}
u_-=T_{5}(\l_*,a_-)f_--c_-\psi_*=T_{5}(\l_*,a_-)f_--\psi_*.
\end{equation*}
The relations obtained in this way together with (\ref{4.2})
lead to (\ref{1.13}).
\end{proof}

\begin{proof}[Proof of Theorem~\ref{th1.3}]
The general lines of the proof are similar to those of the
previous one. According to the results of the previous section the
eigenvalues of $H$ converging to $\l_*$ are roots of the equation
(\ref{4.17}). First we will check that the function at the
left-hand side of this equation has either two simple zeroes or
one second-order zero converging to $\l_*$ as $l\to+\infty$.

To this aim, we need to estimate the functions $A_{ij}$.
Proposition~\ref{lm1.2} implies
\begin{equation}\label{5.7}
\phi_m^-(x_1,l,a_+)=(-1)^m\tau
c(\l_*,a_+)\E^{-2\kappa_1^+(\l_*)l}\E^{-\kappa_1^+(\l_*)x_1}+
\Odr(\E^{-2 \rho(\l_*) l}),
\end{equation}
This formula together with (\ref{5.1}) allow us to conclude that
\begin{equation}\label{5.8}
A_{ij}(\l,l)=\Odr(\E^{-2\kappa_1^+(\l_*)l}),
\end{equation}
hence for any small $\d$ we have the inequality
\begin{equation*}
|(\l-\l_*)\tr A(\l,l)+\det A(\l,l)|<\d^2\quad \text{as}\quad
|\l-\l_*|=\d,
\end{equation*}
if $l$ is large enough. Since the functions $A_{ij}$ are
holomorphic, by Rouch\'{e} theorem this inequality implies that
the function $\l\mapsto D(\l,l):=\det\big((\l-\l_*)
\mathrm{E}+\mathrm{A}(\l,l)\big)$ has the same number of zeroes
(with the order taken into account) as the function
$\l\mapsto(\l-\l_*)^2$ does. The last function has $\l_*$ as a
second-order zero, of course, so it follows that the function
$D(\cdot,l)$ has either two simple zeroes or a second-order zero,
converging to $\l_*$ as $l\to+\infty$. In what follows we denote
these roots as $\l^\pm$, the case of the second-order zero
corresponds to the equality $\l^+=\l^-$.

As it was established in the previous section, the nontrivial
solutions of the equation (\ref{3.4a}) associated with the roots
of (\ref{4.17}) are given by (\ref{4.17a}) with the coefficients
$c_\pm$ solving the system of linear equations (\ref{4.16}). If
the numbers $\l^\pm$ solve (\ref{4.20a}), the system (\ref{4.16})
has at least one nontrivial solution corresponding to $\l^+$ and
$\l^-$.

Suppose that $\l^+\not=\l^-$. Then $\l^\pm$ are simple zeroes of
the function $D(\cdot,l)$, and in view of the above discussion
the system (\ref{4.16}) has exactly one non-trivial solution for
$\l=\l^+$ and $\l=\l^-$. Hence in the case $\l^+\not=\l^-$ the
operator $H$ has exactly two simple eigenvalues converging to
$\l_*$ as $l\to+\infty$.

Let us check that if the system (\ref{4.16}) has two linear
independent solutions referring to $\l=\l^\pm$ it follows that
$\l^\pm$ is a second-order zero of the function $D(\cdot,l)$.
Indeed, two linear independent solutions exist if and only if
\begin{equation}\label{5.19}
A_{11}(\l^\pm,l)=A_{22}(\l^\pm,l)=\l_*-\l^\pm,\quad
A_{12}(\l^\pm,l)=A_{21}(\l^\pm,l)=0.
\end{equation}
The derivative of $D(\l,l)$ with respect to $\l$ equals
\begin{align*}
\frac{\p D}{\p\l}(\l,l)=&2(\l-\l_*)+(A_{11}(\l,l)+A_{22}(\l,l))
\\
&+(\l-\l_*)\left(\frac{\p A_{11}}{\p\l}(\l,l)+\frac{\p
A_{22}}{\p\l}(\l,l)\right)
\\
&+A_{11}(\l,l)\frac{\p A_{22}}{\p\l}(\l,l)-A_{12}(\l,l)\frac{\p
A_{21}}{\p\l}(\l,l)
\\
&+A_{22}(\l,l)\frac{\p A_{11}}{\p\l}(\l,l)-A_{21}(\l,l)\frac{\p
A_{12}}{\p\l}(\l,l).
\end{align*}
Substituting from (\ref{5.19}) into this expression, we see that
\begin{equation*}
\frac{\p D }{\p\l}(\l,l)=0 \quad\text{as}\quad \l=\l^\pm,
\end{equation*}
thus $\l^\pm$ is a second-order zero.

It is more complicated to check existence of a double eigenvalue
of the operator $H$ if $\l^+=\l^-=:\widetilde{\l}$. It is
equivalent to the fact that for $\l=\widetilde{\l}$ the system
(\ref{4.16}) has two linear independent solutions, and this in
turn is equivalent to the relations (\ref{5.19}). Let us prove
that they hold. Consider the boundary value problem
\begin{equation}\label{3.55}
\begin{gathered}
(\D+\l)u=0,\quad x\in\Pi_a\setminus(\g_+\cup\g_-), \qquad
u=0,\quad x\in\p\Pi,
\\
\frac{\p u}{\p x_2}\Big|_{x_2=+0}- \frac{\p u}{\p
x_2}\Big|_{x_2=-0}=-g_\pm, \quad x\in\g_\pm.
\end{gathered}
\end{equation}
Here $g_\pm\in L_2(\g_\pm)$ are arbitrary functions, and the
parameter $\l$ is supposed to range in a small neighborhood of
$\l_*$ without coinciding with $\l_*$ and $\widetilde{\l}$. This
problem is uniquely solvable provided we seek a
$L_2(\Pi)$-solution to (\ref{3.55}). In a complete analogy with
the proof of Lemma~\ref{lm4.1} one can check easily that the
problem (\ref{3.55}) is equivalent to the equation
\begin{equation}\label{3.56}
\boldsymbol{f}+T_{8}(\l,l,\boldsymbol{a})
\boldsymbol{f}=\boldsymbol{g},
\end{equation}
where $\boldsymbol{g}=(g_+,g_-)\in L_2(\g_{a_-})\oplus
L_2(\g_{a_+})$, while the solution $u$ of (\ref{3.55}) is given by
\begin{equation*}
u(x,\l,l)=u_+(x_1-l,x_2,\l,l)+u_-(x_1+l,x_2,\l,l),\quad
u_\pm:=T_{3}(\l,a_\pm)f_\pm.
\end{equation*}
We can solve the equation (\ref{3.56}) in the same way as the
equation (\ref{4.3a}), obtaining as a result that
\begin{equation}\label{3.58}
\boldsymbol{f}+\frac{T_{4}^+
\boldsymbol{f}}{\l-\l_*}\boldsymbol{\Phi}_*^+ +
\frac{T_{4}^-\boldsymbol{f}}{\l-\l_*}\boldsymbol{\Phi}_*^-=
\boldsymbol{G},\quad \boldsymbol{G}:=
\big(\I+T_{9}(\l,l,\boldsymbol{a})\big)^{-1}\boldsymbol{g}.
\end{equation}
Hence the function $\boldsymbol{f}$ is of the form
\begin{equation}\label{3.59}
\boldsymbol{f}=C_+\boldsymbol{\Phi}^+_*+
C_-\boldsymbol{\Phi}^-_*+\boldsymbol{G},
\end{equation}
where $C_\pm=C_\pm(\l,l)$ are constants to be found. Denoting
$\boldsymbol{C}:=\left(\begin{smallmatrix} C_+ \\ C_-
\end{smallmatrix}\right)$ and substituting (\ref{3.59}) into
(\ref{3.58}), we obtain an equation for $\boldsymbol{C}$,
\begin{equation}\label{3.60}
\big((\l-\l_*)\mathrm{E}+\mathrm{A}(\l,l)\big) \boldsymbol{C}
=\boldsymbol{h},\quad \boldsymbol{h}:=
\begin{pmatrix}
-T_{4}^+ \boldsymbol{G}
\\
-T_{4}^- \boldsymbol{G}
\end{pmatrix}.
\end{equation}
The solution of this system is given by Cramer's formula,
\begin{equation}\label{3.61}
\begin{aligned}
&C_+(\l,l)=\frac{A_{12}(\l,l)T_{4}^-\boldsymbol{G}-
\big(\l-\l_*+A_{22}(\l,l)\big) T_{4}^+\boldsymbol{G}}
{D(\l,l)}\,,
\\
&C_-(\l,l)=\frac{A_{21}(\l,l)
T_{4}^+\boldsymbol{G}-\big(\l-\l_*+
A_{11}(\l,l)\big)T_{4}^-\boldsymbol{G}}{D(\l,l)}\,.
\end{aligned}
\end{equation}
Using now (\ref{3.60}) and Lemma~\ref{lm3.13}, we infer that
\begin{equation}\label{3.62}
\begin{aligned}
u_+(\cdot,\l,l)=&-C_+(\l,l)\psi_m(\cdot,a_+)+C_+(\l,l)T_{5}(\l,a_+)
\Phi_{*,+}^+
\\
&+ C_-(\l,l)T_{5}(\l,a_+) \Phi_{*,+}^-+ T_{5}(\l,a_+)G_+,
\\
u_-(\cdot,\l,l)=&-C_-(\l,l)\psi_n(\cdot,a_-)+C_-(\l,l)T_{5}(\l,a_-)
\Phi_{*,-}^+
\\
&+ C_-(\l,l)T_{5}(\l,a_-)\Phi_{*,-}^-+ T_{5}(\l,a_-)G_-,
\end{aligned}
\end{equation}
where $\Phi_{*,\pm}^\pm$ and $G_\pm$ are the components of the
vectors $\boldsymbol{\Phi}_*^\pm$ and $\boldsymbol{G}$,
\begin{equation*}
\boldsymbol{\Phi}_*^\pm=(\Phi_{*,+}^\pm,\Phi_{*,-}^\pm),\quad
\boldsymbol{G}=(G_+,G_-).
\end{equation*}
Since the number $\widetilde{\l}$ is a second-order zero of
$D(\cdot,l)$, we conclude from (\ref{3.61}) that the coefficients
$C_\pm$ have, in general, a second-order pole at $\widetilde{\l}$,
and the same is true for $u_\pm$. Taking into account
(\ref{3.62}), we conclude that the solution of (\ref{3.55}) can be
represented as
\begin{equation}\label{3.63}
u(x,\l,l)=u_{-1}^+(x,\l,l)C_+(\l,l)+ u_{-1}^-(x,\l,l)C_-(\l,l)
+\Odr(1),\quad \l\to\widetilde{\l}.
\end{equation}
In a complete analogy with the proof of Lemma~\ref{lm3.13} one
can check easily that the solution of the problem (\ref{3.55})
has a simple pole at $\widetilde{\l}$. Hence the function
$u_{-1}^+(x,\l,l)C_+(\l,l)+ u_{-1}^-(x,\l,l)C_-(\l,l)$  has a
simple pole at $\widetilde{\l}$. For $x$ from a neighborhood of
$\g_+$ this function satisfies due to (\ref{3.61}), (\ref{3.62})
the relation
\begin{align*}
&D(\l,l)\left(u_{-1}^+(x,\l,l)C_+(\l,l)+
u_{-1}^-(x,\l,l)C_-(\l,l)\right)
\\
&= \Big(\big(\l-\l_*+A_{22}(\l,l)\big)T_{4}^+\boldsymbol{G}-
A_{12}(\l,l)T_{4}^-\boldsymbol{G} \Big)\psi_m^+(x_1-l,x_2)+
\Odr(\E^{-2\kappa_1^+(\l_*)l}).
\end{align*}
Since $\widetilde{\l}$ is by assumption a second-order zero of
$D(\cdot,l)$, the obtained identity yields that
\begin{equation*}
\widetilde{\l}-\l_*+A_{22}(\widetilde{\l},l)=
A_{12}(\widetilde{\l},l)=0.
\end{equation*}
Observing the behavior of the function $u$ for $x$ in the
vicinity of $\g_-$, one can prove in the same way that
\begin{equation*}
\widetilde{\l}-\l_*+A_{11}(\widetilde{\l},l)=
A_{21}(\widetilde{\l},l)=0.
\end{equation*}
This completes the check of the relations (\ref{5.19}) for
$\l^+=\l^-$ showing that in this case the operator $H$ has a
double eigenvalue converging to $\l_*$ as $l\to+\infty$.

We proceed to calculation of the asymptotic expansions for the
root(s) of the equation (\ref{4.17}). Substituting the estimates
(\ref{5.8}) into (\ref{4.17}) we obtain
\begin{equation}\label{5.9}
\l-\l_*=o(\E^{-\kappa_1^+(\l_*)l}).
\end{equation}
This relation in combination with (\ref{5.1}), (\ref{5.4a}) and
the estimate (\ref{3.21a}) imply that
\begin{equation}\label{5.10}
A_{22}(\l,l)=\Odr(l\E^{-4\kappa_1^+(\l_*)l}).
\end{equation}
It is easy to establish an expression for $A_{11}$ similar to
(\ref{5.4a}), which together with (\ref{5.7}) and (\ref{5.9})
yield
\begin{equation}\label{5.11}
A_{11}(\l,l)=\Odr(l\E^{-4\kappa_1^+(\l_*)l}).
\end{equation}
Proceeding in the same way as in (\ref{5.4a}) we obtain a chain of
relations,
\begin{align*}
A_{12}(\l,l)&=T_{4}^+(\I+T_{9}(\l,l,\boldsymbol{a}))^{-1}
\boldsymbol{\phi}_*^-(\cdot,l)=
T_{4}^+\boldsymbol{\phi}_*^-(\cdot,l)+
\Odr(\|T_{9}\|\|\boldsymbol{\phi}_*^-\|)
\\
&=(\phi_n^+(\cdot,l,a_-),\psi_m(\cdot,a_+))_{L_2(\g_{a_+})}+
\Odr(l\E^{-4\kappa_1^+(\l_*)l}).
\end{align*}
Due to (\ref{5.1}) and (\ref{1.5}) we have
\begin{align*}
(\phi_n^+(\cdot,l,a_-),&\psi_m(\cdot,a_+))_{L_2(\g_{a_+})}
\\
&= c(\l_*,a_-)\E^{-2\kappa_1^+(\l_*)l}
\big(\E^{-\kappa_1^+(\l_*)x_1},\psi_m(\cdot,a_+)\big)_{L_2(\g_{a_+})}+
\Odr(\E^{-2\rho(\l_*)l})
\\
&=\mu(l,\boldsymbol{a})\E^{-2\kappa_1^+(\l_*)l}+
\Odr(\E^{-2\rho(\l_*)l}),
\end{align*}
where $\mu(l,\boldsymbol{a})$ is given by (\ref{1.15}).
Consequently,
\begin{equation}\label{5.12}
A_{12}(\l,l)=\mu(l,\boldsymbol{a}) \E^{-2\kappa_1^+(\l_*)l}+
\Odr(\E^{-2\rho(\l_*)l}+l\E^{-4\kappa_1^+(\l_*)l}),
\end{equation}
and in the same way one can show that
\begin{equation}\label{5.13}
A_{21}(\l,l)=\mu(l,\boldsymbol{a}) \E^{-2\kappa_1^+(\l_*)l}+
\Odr(\E^{-2\rho(\l_*)l}+l\E^{-4\kappa_1^+(\l_*)l}).
\end{equation}
The equation (\ref{4.17}) is equivalent to the following pair of
the equations,
\begin{equation}\label{5.14}
\l-\l_* =\frac{-\tr A(\l,l)\pm\sqrt{(A_{11}(\l,l)-A_{22}(\l,l))^2+
4A_{12}(\l,l)A_{21}(\l,l)}}{2}\,.
\end{equation}
If $c(\l_*,a_-)c(\l_*,a_+)=0$, these equations together with
(\ref{5.10})--(\ref{5.13}) imply that
\begin{equation*}
\l-\l_*=\Odr(\E^{-2\rho(\l_*)l}+l\E^{-4\kappa_1^+(\l_*)l}).
\end{equation*}
which proves the asymptotic expansion (\ref{1.14}) in the case
$\mu(l,\boldsymbol{a})=0$.

Suppose on the contrary that $c(\l_*,a_-)c(\l_*,a_+)\not=0$. In
this case the function $(A_{11}-A_{22})^2+4A_{12}A_{21}$ is
non-zero as $\l=\l_*$, and therefore its square root is
holomorphic w.r.t. $\l$. Using this fact and the relations
(\ref{5.10})--(\ref{5.13}), one can show easily in analogy with
the similar argument for the equation (\ref{4.17}) that each of
the equations (\ref{5.14}) has a unique root converging to
$\l_*$ as $l\to+\infty$. Hence one of the roots of (\ref{4.17})
satisfies the first of the equations (\ref{5.14}), while the
other satisfies the other one. Substituting now from
(\ref{5.10})--(\ref{5.13}) into (\ref{5.14}), we arrive
immediately at the asymptotics (\ref{1.14}), (\ref{1.15}) in the
case $\mu(l,\boldsymbol{a})\not=0$.
\end{proof}

\begin{proof}[Proof of Theorem~\ref{th1.4}]
Let $\boldsymbol{c}$ be a nontrivial solution to the system
(\ref{4.16}), where $\l$ is $\l^+$ or $\l^-$. Without loss of
generality we may assume that $\|\boldsymbol{c}
\|_{\mathbb{R}^2}=1$. Modifying (\ref{4.17a}), we choose the
corresponding nontrivial solution of the equation (\ref{3.4a}) as
$\boldsymbol{f}=-c_+\boldsymbol{\Phi}^+_*- c_-
\boldsymbol{\Phi}^-_*$. In analogy with (\ref{5.9a}) we then
obtain
\begin{align*}
u_-&=\frac{\psi_n(\cdot,a_-)}{\l-\l_*}T_{4}^-\boldsymbol{f}
+\Odr(\E^{-2\kappa_1^+(\l_*)l})
\\
&=-\frac{c_+A_{21}(\l,l)+c_-A_{22}(\l,l)}
{\l-\l_*}\psi_n(\cdot,a_-)+\Odr(\E^{-2\kappa_1^+(\l_*)l}),
\end{align*}
which holds true in $\H^1(\Pi_{a_-})$ and $\H^2(S)$ for each
$S\in\Xi_{a_-}$. Employing now the system (\ref{4.16}) we can
write
\begin{equation*}
c_+A_{21}(\l,l)+c_-A_{22}(\l,l)=-c_-(\l-\l_*),
\end{equation*}
hence
\begin{equation*}
u_-=c_-\psi_n(\cdot,a_-)+\Odr(\E^{-2\kappa_1^+(\l_*)l}),
\end{equation*}
and in the same way one can prove that
\begin{equation*}
u_+=c_+\psi_m(\cdot,a_+)+\Odr(\E^{-2\kappa_1^+(\l_*)l}).
\end{equation*}
in the norm of $\H^1(\Pi_{a_-})$ and $\H^2(S)$ for each $S\in
\Xi_{a_-}$. The last two relations prove the sought formul{\ae}
(\ref{1.18}).

Suppose that $\l^+=\l^-$, then (\ref{4.16}) has two nontrivial
solutions, which means that $(\l-\l_*)\mathrm{E}
+\mathrm{A}(\l,l)=0$; we can choose these solutions as
$(c_+,c_-)=(-1,0)$ and $(c_+,c_-)=(0,-1)$. Substituting these
values into (\ref{1.18}), we arrive at (\ref{1.17}).

Suppose that $\mu(l,\boldsymbol{a})\not=0$. In view of
(\ref{1.14}) it implies that $\l^+(\l,a)\not=\l^-(\l,a)$, i.e.
that $\l^\pm(\l,a)$ are simple eigenvalues. In this case the
relations (\ref{1.14}) and (\ref{5.11}), (\ref{5.12}) yield
\begin{equation}\label{5.21}
\begin{aligned}
&\l^\pm-\l_*+A_{11}(\l^\pm,l)=\pm|\mu(l,\boldsymbol{a})|\E^{-2\kappa_1^+(\l_*)l}
(1+\Odr(\E^{-2\kappa_1^+(\l_*)l}))\not=0,
\\
&A_{12}(\l^\pm,l)=\mu(l,\boldsymbol{a})\E^{-2\kappa_1^+(\l_*)l}(1+
\Odr(\E^{-2\kappa_1^+(\l_*)l}))\not=0.
\end{aligned}
\end{equation}
Since the matrix $(\l^\pm-\l_*)\mathrm{E}+\mathrm{A}(\l^\pm,l)$
has rank one, we can choose nontrivial solutions of (\ref{4.16})
as
\begin{align*}
&c_+^\pm:=\pm\frac{\sqrt{2}(\l^\pm-\l_*+A_{11}(\l^\pm,l))}
{\sqrt{(\l^\pm-\l_*+A_{11}(\l^\pm,l)^2+A_{12}^2(\l^\pm,l)}}\,,
\\
&c_-^\pm:=\pm\frac{\sqrt{2}A_{12}(\l^\pm,l)}
{\sqrt{(\l^\pm-\l_*+A_{11}(\l^\pm,l)^2+A_{12}^2(\l^\pm,l)}}\,.
\end{align*}
In view of to (\ref{5.21}) we then have
\begin{equation*}
c_+^\pm=1+\Odr(\E^{-2\kappa_1^+(\l_*)l}),\quad
c_-^\pm=\mp\sgn\mu(l,\boldsymbol{a})
+\Odr(\E^{-2\kappa_1^+(\l_*)l}).
\end{equation*}
Substituting from here into (\ref{1.18}) we arrive immediately at
(\ref{1.16}).
\end{proof}

\subsection*{Acknowledgment}

D.B. was supported by \emph{Marie Curie International Fellowship}
within the 6th European Community Framework Programm
(MIF1-CT-2005-006254), and in part by the Russian Foundation for
Basic Researches (No. 06-01-00138). P.E. was supported in part by
the Czech Academy of Sciences and Ministry of Education, Youth and
Sports within the projects A100480501 and LC06002.


\renewcommand{\refname}

\end{document}